\documentclass[authoryear,10pt]{elsarticle}

\usepackage{amssymb}
\usepackage{amsmath}
\usepackage{lineno}
\usepackage{graphicx}
\usepackage{textcomp}
\usepackage{color}
\usepackage{esint} 
\usepackage{hyperref} 

\journal{Advances in Geophysics}

\begin{document}
\begin{frontmatter}
\title{Lava Tube Exploration with LunarLeaper}

\author{Anna Mittelholz$^{a}$, Simon C. Stähler$^{a,b}$, Valentin T. Bickel$^{c}$, Jordan Aaron$^{a}$, Aurelie Cocheril$^{b}$,  Alessandro Ghirotto$^{a}$, Matthias Grott$^{d}$, Svein-Erik Hamran$^{e}$, Ozgur Karaketin$^{f}$, Alessandro Maturilli$^{d}$, Birgit Ritter$^{f}$, Alexis Shakas$^{a}$} %

\affiliation[a]{organization={Department of Earth and Planetary Sciences, ETH Zurich},%
            country={Switzerland}}
            
\affiliation[b]{organization={Space Systems and Technology, ETH Zurich},
            country={Switzerland}}
            
\affiliation[c]{organization={Center for Space and Habitability, University of Bern},%
            country={Switzerland}}

\affiliation[d]{organization={Institute of Planetary Research, German Aerospace Center},%
            country={Germany}}

\affiliation[e]{organization={Department of Technology Systems, University of Oslo}, country={Norway}}

\affiliation[f]{organization={Royal Observatory of Belgium}, country={Belgium}}

\begin{abstract}

Lunar pits, some of which are interpreted as collapse features into underlying lava tubes, expose otherwise inaccessible stratigraphy and may provide entry points to subsurface voids that preserve records of lunar volcanism and offer potential sites for future human exploration. We synthesize the current state of knowledge on lunar pits and lava tubes, covering their morphological characteristics, classification, proposed formation mechanisms, mechanical stability, and detection from orbit. We then review the open science questions that pit and pit-wall investigation is uniquely placed to address, spanning the volcanic stratigraphy of the lunar maria, the structure and lateral variability of the regolith, and the dimensions and accessibility of subsurface conduits. 
To evaluate how these questions can be tackled in situ, we assess the feasibility and expected performance of geophysical and remote-sensing investigations for subsurface voids and surface exposures, mainly focusing on gravity measurements, ground-penetrating radar, high-resolution imaging, and  spectroscopy. Building on this, we present LunarLeaper, a small legged-robot mission concept combining a gravimeter, ground-penetrating radar, high-resolution imager, spectrometer, and leg-based geomechanical experiments to deliver the first in situ investigation of a mare pit. The concept targets the Marius Hills Pit and its associated rille, with a mobility architecture optimized for the rugged terrain encountered at pit edges and funnel slopes.

\end{abstract}

\begin{highlights}
\item We review the current state of knowledge on lunar pits and lava tubes, including their formation, morphology, stability, and detection from orbit.
\item We synthesize the open science questions that pit and pit-wall investigations can address, including mare volcanic stratigraphy, regolith and paleo-regolith structure, and the geometry of subsurface conduits.
\item We evaluate the feasibility and performance of in situ geophysical and remote-sensing surveys, focusing on gravity, ground-penetrating radar, imaging, and spectroscopy, to directly detect and characterize subsurface voids and surface-visible structures during a first lunar pit exploration mission.
\item We present LunarLeaper, a legged-robot mission concept for the first in situ exploration of a mare pit, outlining its mobility system, instruments, and expected scientific return.
\end{highlights}


\end{frontmatter}

\section{Introduction}
\label{Introduction}

The past two decades of space exploration have shown that lava tubes are likely ubiquitous features across terrestrial planets and moons \citep{cushing2007, haruyama_possible_2009, sauro_lava_2020}. While terrestrial lava tubes have been studied in some detail \citep{wang_enhanced_2025, torrese_detection_2021, torrese_planetary_2022, sauro_volcanic_2019}, their extraterrestrial presence is usually inferred indirectly by remote sensing data or morphological signatures known from Earth analogues, such as sinuous rilles (i.e., elongated topographic depressions \citep{cruikshank_lunar_1972, greeley_lava_1971}) and/or skylights (i.e., collapsed cave roofs). Lava tubes form when the surface of a lava flow solidifies while the still-liquid interior continues to drain laterally, leaving behind quasi-horizontal, hollow tunnels. Lunar reduced gravity and high lava discharge rates during the early volcanic history likely enabled the formation and preservation of more extensive subsurface systems than those on Earth \citep{sauro_lava_2020}. 

\begin{figure}
    \centering
    \includegraphics[width=1\linewidth]{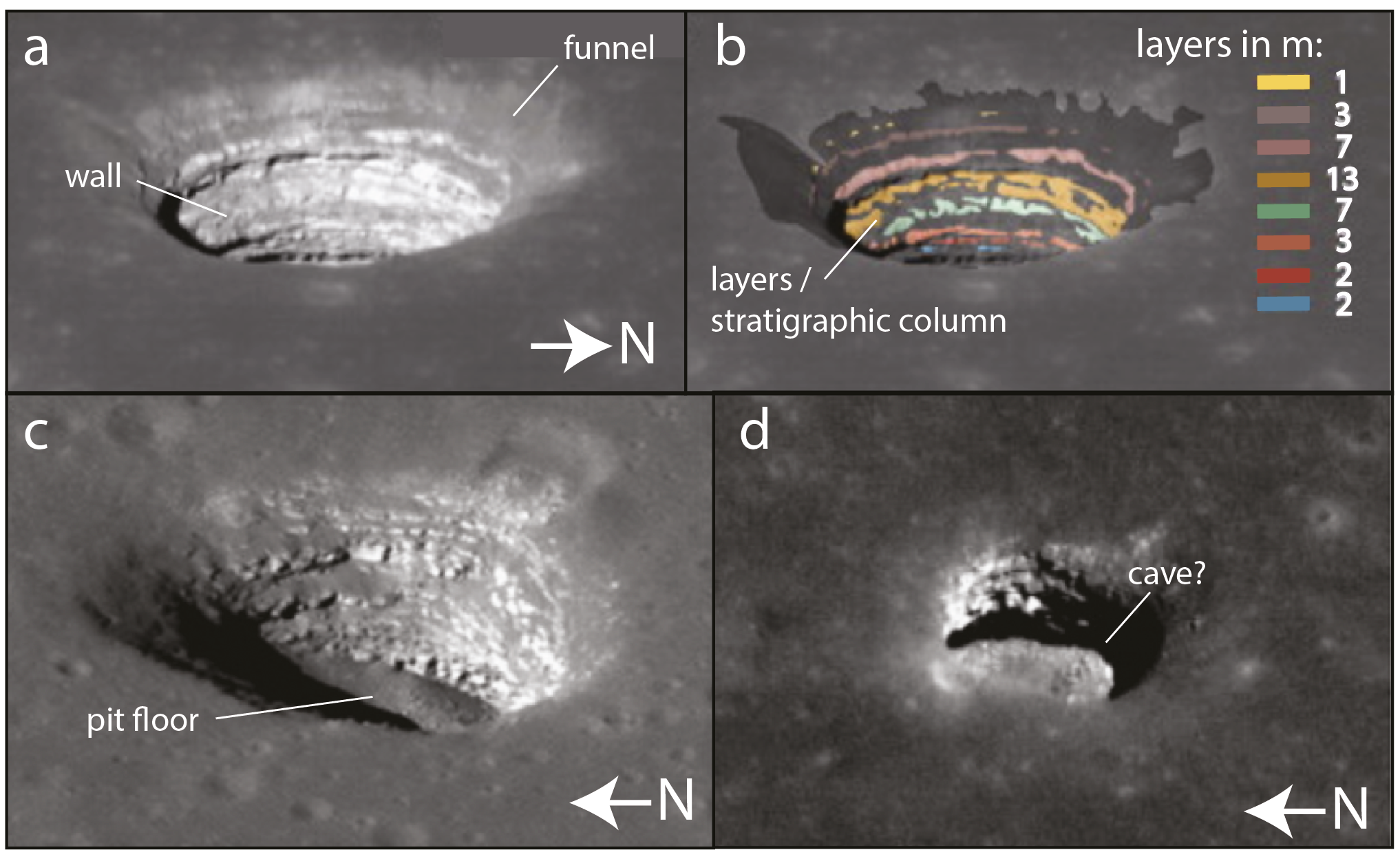}
    \caption{Oblique orbital images taken of exposed pit walls of (a+b) Mare Tranquillitatis Pit, (c) Mare Ingenii Pit and (d) Marius Hills Pit; key morphological characteristics of pits are highlighted. (b) shows mapping of individual layers and their approximate thickness in meters. Modified from \cite{robinson_confirmation_2012}. }
    \label{pitwall}
\end{figure}

\begin{figure}
    \centering
    \includegraphics[width=0.9\linewidth]{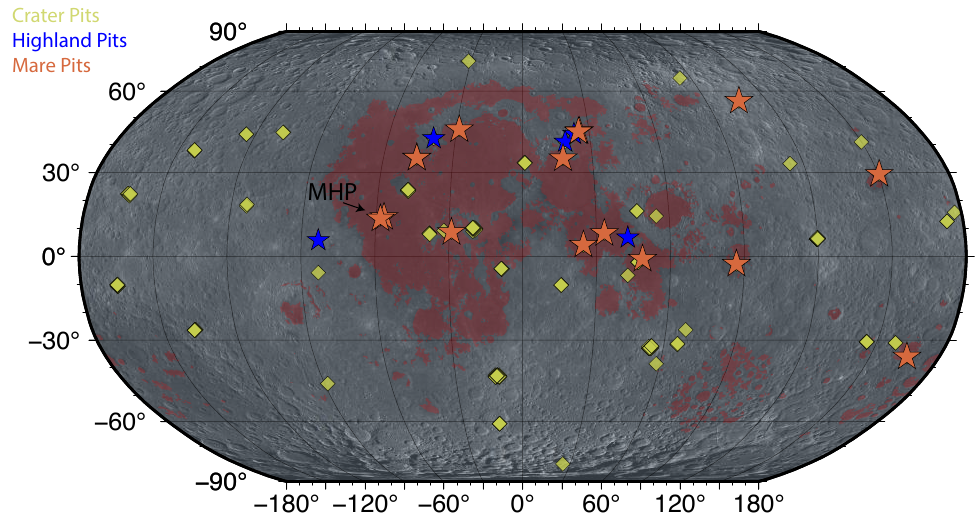}
    \caption{The global distribution of known pits, a total of 279 pits from \cite{m_s_wagner_r_v_occurrence_2021} with the updated selection of 5 highland and 16 mare pits mapped by \cite{wagner_lunar_2022}, plotted on a global LRO WAC basemap; mare regions are highlighted in red. Note that many pits are directly co-located, leading to overlapping markers on the map. }
    \label{pit_distribution}
\end{figure}

While the existence of lunar lava tubes had previously been proposed \citep{greeley_lava_1971, murase_viscosity_1970}, the first confirmation of this hypothesis was provided by the detection of the Marius Hills Pit (MHP) in KAGUYA/Selene orbital imagery (\cite{haruyama_possible_2009}; Figure \ref{pitwall}d); since then, approximately 300 lunar pits have been mapped across the lunar surface (Figure \ref{pit_distribution}), predominantly in Lunar Reconnaissance Orbiter (LRO) Near Angle and Wide Angle Camera (NAC and WAC) data \citep{wagner_distribution_2014, wagner_lunar_2022}. Pits are generally assumed to be collapse structures into pre-existing subsurface voids such as lava tubes, which is why they are also called skylights (\cite{robinson_confirmation_2012, xu_lunar_2024}; Figure \ref{pitwall}). Typically, pits feature a central, collapsed section, often with near-vertical walls of exposed volcanic deposits, surrounded by a funnel. The funnel is usually covered by regolith and potentially meter- to sub-meter-sized boulders. The topographic slope of the funnel is likely constrained by the angle of repose of dry regolith, at about 30$^{\circ}$ to 40$^{\circ}$. A more detailed exploration of pits and pit walls requires an asset to traverse the funnel.

Different classes of pits have been identified (Figure \ref{pit_distribution}): Mare, crater and highland pits. 'mare' and 'highland' describe the location of the pit (the geologic territory), whereas 'crater' pits (also called 'impact melt pits') are formed as the result of high energy dissipation during an impact heating the surface, which then drains, cools, and creates voids due to deformation stresses. 
While all of those pits offer access to the local subsurface, mare pits are most likely to be of volcanic origin and lead to more extensive, larger-scale tube systems. Overall, 16 mare pits have been identified in orbital images so far; these mare pits are the focus of this review \citep{wagner_lunar_2022}. However, orbital imaging alone provides only indirect evidence of subsurface voids, as it is largely limited to surface morphology and cannot directly resolve the geometry, extent, or continuity of underlying lava tube systems.

Orbital geophysical datasets also provide important but indirect constraints on the presence and geometry of subsurface voids associated with lunar lava tubes. Gravity measurements from the GRAIL mission reveal localized mass deficits and linear gravity anomalies that are consistent with tube-scale voids buried beneath mare basalts \citep{zhu_grail_2024, chappaz_evidence_2017}. However, these signals represent spatial averages over tens of kilometers, limiting the ability to distinguish individual tubes or to resolve their precise geometry. As such, identified mass deficits within the Marius Hills region do not directly underlay the MHP, the region in which a lava tube is expected, but are an indicator for possible tube systems in the broader area. 
In addition, complementary orbital radar sounding has revealed dielectric contrasts and subsurface reflectors that may indicate buried cavities or coherent lava-tube segments \citep{carrer_radar_2024,kaku_detection_2017}. Radar penetration depth and horizontal resolution significantly degrade in dense basaltic terrains, and clutter from (near-)surface roughness can obscure or mimic true subsurface features \citep[e.g.,][]{lai_first_2020}. As a result, while orbital gravity and radar provide compelling evidence for extensive subsurface void systems, their spatial resolution remains insufficient for unambiguous detection, characterization, or validation without targeted in situ measurements. 

Lunar pits and their anticipated lava tubes have become focal points of scientific and exploration interest. 
This is because the accessible stratigraphy exposed along a pit wall offers unique insights into lunar volcanic history. 
Lava tubes may provide access to undisturbed, ancient volcanic deposits and - in polar regions - to volatile species \citep{lee2018possible}, providing insights into the Moon’s interior and thermal evolution that is otherwise destroyed on the heavily modified surface. 
Pits provide openings into these environments, enabling direct robotic or crewed access for mapping, resource assessment, and environmental characterization. 
In addition, subsurface voids can provide natural shielding from radiation, micrometeorite bombardment, and extreme surface temperature variations \citep{angelis_lunar_2002, horvath_thermal_2022}. These characteristics make lunar caves compelling candidates for future habitats and long-duration science stations, as long as some of the challenges of accessing subsurface voids can be reliably addressed; this includes safe descent/ascent and potential structural instability like roof collapse.
As a result, lunar pits are increasingly prioritized as targets for precursor missions aimed at evaluating their geometry, stability, accessibility, and suitability for scientific investigation and in situ resource utilization. 

One such mission concept is LunarLeaper, designed to perform the first in situ exploration of a mare pit and its associated subsurface voids \citep{mittelholz_lunarleaper_2024,kolvenbach_lunarleapermission_2026}. LunarLeaper relies on legged locomotion \citep{arm_scientific_2023}, allowing agile traversing across uneven and steep terrain, as one would encounter at the pit. Equipped with a ground-penetrating radar, gravimeter, imager, and spectrometer, LunarLeaper aims to detect and map subsurface lava tubes while simultaneously characterizing the surrounding surface terrain, including the pit funnel, walls, and floor. 
This mission would provide the first surface measurements required to validate remote-sensing interpretations, further constrain models of lunar volcanic evolution, and evaluate the potential of subsurface voids for shelter and habitation.

In this paper, we lay out the importance of studying lunar lava tubes in the context of some of the major outstanding questions in lunar volcanic evolution. 
We begin by reviewing the geographic distribution of all known mare pits, including their geological setting and morphology, and discuss what these features reveal about some of the key open questions regarding lunar volcanic evolution (Section 2). 
Understanding the formation and mechanical stability of these pits and any associated lava tubes as discussed in Section 3 is crucial to interpret their volcanic origin and to assess their suitability as targets for future exploration.
We then review a suite of geophysical techniques capable of detecting and characterizing lava tubes from orbit and the surface, showcasing the LunarLeaper instrument suite and some initial investigations exploring their sensitivities to relevant parameters. 
Finally, we describe the LunarLeaper mission concept as a promising approach for in situ investigation of lunar pits and their potential connection to extensive subsurface voids. The mission is designed to provide the first detailed reconnaissance of a mare pit, enabling both scientific discovery and the validation of technologies required for future subsurface exploration.

\section{Studying Lunar Volcanic Evolution at and around Lunar Pits}
\label{volc_evo}

The walls of lunar mare pits expose a detailed stratigraphic record of millions of years of volcanic evolution that would otherwise be inaccessible, buried beneath lava flows. Combined with regional geophysical mapping, observations obtained within and around pits allow reconstruction of both local and regional geological evolution. 

Lunar mare pits are thought to be secondary collapse features that formed long after the emplacement of the mare lava flows, when the roofs of subsurface voids (likely former lava tubes) became unstable and collapsed. The MHP formed in maria older than 3.3 Ga (Ga is $1\times10^9$ years) and \cite{wagner_distribution_2014} argue that it could not have been preserved in its current state had it formed contemporaneously with the surrounding mare. This reasoning applies to mare pits in general, as mare basalts are thought to be older than 2 Ga~\citep{Wilhelms1987}.
Given that a simple crater with a diameter of ~100 m on the Moon is expected to persist for only on the order of 100 Myr~\citep{gault_saturation_1970}, lunar pits are therefore inferred to be comparatively young features. 
Their formation may have been triggered by small-scale impact events, impact-induced seismic shaking, or tectonic activity occurring well after mare emplacement~\citep{martellato_numerical_2013}.

\subsection{Pit Wall Stratigraphy}

Pits provide direct access to the local stratigraphic column, offering information that is otherwise not attainable through orbital remote sensing alone. Previous analyses of LRO NAC images, particularly of the Mare Tranquillitatis Pit, show discrete layers along the wall that appear between 1-13 meters thick (\cite{wagner_distribution_2014}; Figure \ref{pitwall}b). However, orbital observations are not resolved enough to identify sub-meter layering, and interpretations are further constrained by viewing geometry and illumination conditions.

High-resolution, centimeter-scale imaging of pit walls from the surface would allow for significantly improved quantification of mare emplacement processes. Such measurements are essential for constraining flow volumes (i.e., thickness of individual deposits), temporal evolution between eruptions (i.e., presence/absence of paleo-regolith between flow events), and the thermal conditions required for sustained magmatism (i.e., composition of deposits and changes over time/depth).

These parameters are directly relevant to understanding lunar eruption styles, volatile volcanic budgets, transient lunar atmospheres, and the long-term thermal evolution of the lunar interior \citep{shearer_invited_1999, needham_lunar_2017}. Notably, potentially preserved paleo-regolith horizons would mark periods of volcanic quiescence and could retain implanted particles from galactic cosmic rays, solar energetic particle events, and potentially even components of Earth’s early atmosphere \citep{ozima_terrestrial_2005, ozima_toward_2008, fagents_preservation_2010}. Their composition and structure provide critical constraints on the temporal evolution of the lunar surface environment, impact flux, and solar and galactic radiation conditions.

Compositional analyses of exposed lava flow deposits also provide a means of probing the heterogeneity and evolution of the lunar mantle. Mare basalts are typically enriched in $FeO$ and $TiO_2$, depleted in $Al_2O_3$ and are interpreted as products of of remelting of mantle cumulates formed during early differentiation \citep{shearer_invited_1999}. While Apollo and Luna samples suggested a bimodal distribution of $Ti$ content in mare basalts \citep{taylor_planetary_1983}, subsequent orbital observations could not confirm this \citep{giguere_titanium_2000}. Instead, a continuous range of $TiO_2$ content from $<$1 to $\sim$14\% in lunar mare basalt support heterogeneous mantle composition and/or complex mantle dynamics \citep{giguere_titanium_2000}. Here, time dependence is not evident, but also difficult to assess. However, the exposed pit layers provide a unique location to gain understanding of ancient melt sources and magma compositional evolution over time/depth.  

\subsection{Regional Volcanic Evolution}

The broader geological context is particularly significant in regions of mare pits, such as the MHP, which spans $\sim200 \times 250 \text{km}^2$ and exhibits some of the highest densities of volcanic constructs on the Moon. 
Beyond local stratigraphy, the Marius Hills region is among the most volcanically diverse and active regions on the Moon, characterized by some of the thickest known mare basalt sequences \citep{gong_thicknesses_2016} as well as signs of both extrusive and intrusive volcanic activity \citep{kiefer_gravity_2013}. 
An anomalous concentration of granular flow features observed in this region \citep{bickel_global_2022} supports the possibility of recent tectonic or compositional influences. %

\begin{figure}
    \centering
    \includegraphics[width=1\linewidth]{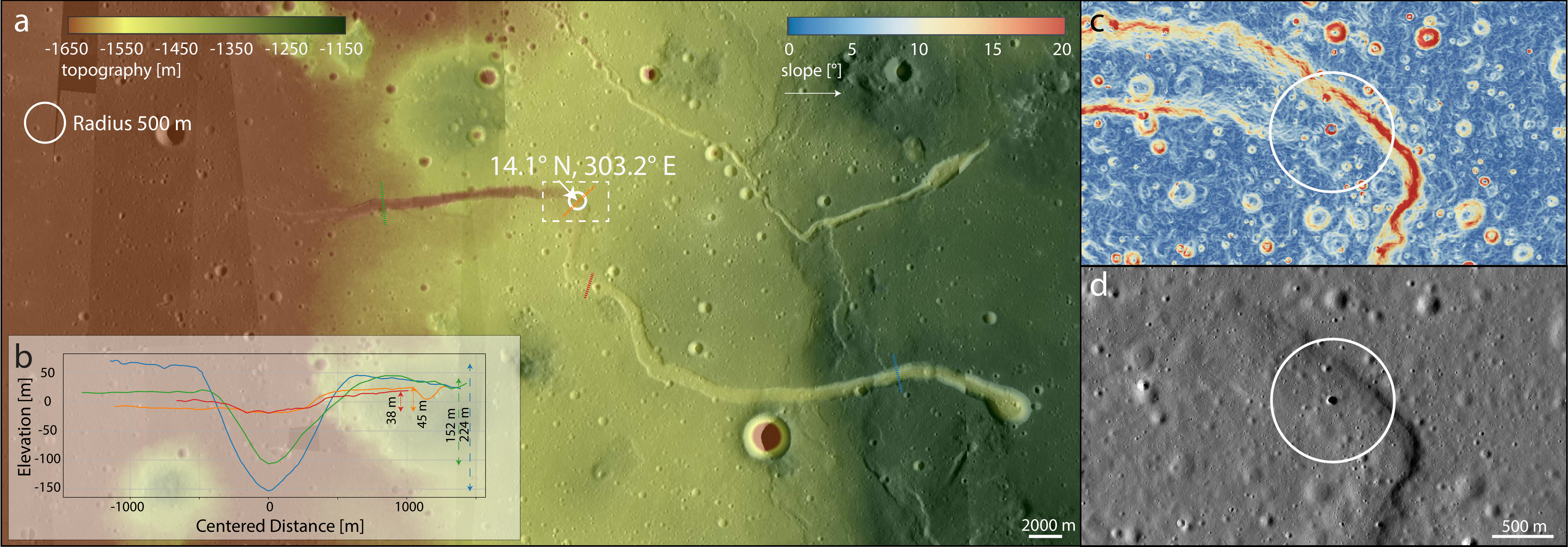}
    \caption{(a) The Marius Hills Pit and its surroundings including the 10s of km long rille structure. The image is a LRO WAC / NAC mosaic overlain by the dynamic Terrain Height DEM from quickmap. Transects across the pit are plotted in (b) according to colors in (a). They show demeaned topography centered at their minimum value. (c+d) Zoom-in on the pit where (c) the slope map highlights the rille structure. (d) LRO WAC / NAC mosaic. Note that the circles in a,c,d show a 500 m radius circle centered on the Marius Hills Pit. }
    \label{MH_Area}
\end{figure}

A common volcanic landform in mare regions, including the Marius Hills, are sinuous rilles, large winding channel-like features that can extend for hundreds of kilometers and reach widths of several kilometers (Figure \ref{MH_Area}). 
A sinuous rille is thought to form when low-viscosity basaltic lava flows across the lunar surface at high effusion rates, carving or constructing a winding channel as it descends along the local slope \citep{hurwitz_lunar_2013, greeley_lava_1971}.  Near its source, the lava may erode deeply into the pre-existing lava plains through thermal, mechanical, or thermo-mechanical processes. Further downslope, as slopes decrease, the flow typically transitions to a more constructional regime in which levees form along the channel margins, confining the lava, and gradually building a shallower, wider channel. While many sinuous rilles appear as continuous surface channels, some exhibit discontinuous surface expressions \citep{Qian2021, Williams2025}. In these cases, the lava flow is interpreted to have transitioned into a subsurface lava tube, with the channel continuing underground across segments where no surface rille is preserved.

In some cases, lava might be transported within insulated lava tubes, and later roof collapse exposes a sinuous depression that mimics a surface channel \citep{hurwitz_lunar_2013}.
These different processes including erosion, levee-bounded channel construction, and (in some cases) collapsed lava tubes, often occur sequentially along the same flow, producing the characteristic downstream evolution from deeply entrenched channels to broad, shallower, more constructional forms. On the Moon, where gravity is low and eruptions were capable of sustaining immense, turbulent flows, these mechanisms operated at scales far exceeding terrestrial analogs, giving rise to the exceptionally wide, deep, and continuous rilles observed today.

The MHP is placed on top of a 10's of km long and approximately 500 m wide rille, a comparatively small rille \citep{qian_longest_2021}, that originates from a source depression at its Easternmost end (Figure \ref{MH_Area}). Near the source region and west of the pit, the rille incises the surrounding terrain by more than 200 m. In contrast, at the MHP location, the relief decreases to only a few tens of meters, before increasing again downstream to >100 m prior to the rille’s disappearance. The origin of this along-rille morphological variation is unknown but may be related to the rille's formation processes itself, to pre-existing topography, or the presence of an uncollapsed lava tube beneath the central section.
Because subsurface data at the spatial scale of rilles are currently unavailable, interpretations rely solely on surface morphology and terrestrial analogues. As a result, the internal structure, continuity, and morphology of rille-associated lava tubes remain inferred rather than directly observed.

\subsection{The Lunar Regolith}
Beyond the investigation of volcanic stratigraphy, composition, and evolution, lunar pits and their funnels provide a valuable opportunity to constrain the thickness and geomechanical properties of the lunar regolith, which is key for understanding lunar surface evolution, predicting trafficability and designing infrastructure \citep{carrier_d_w_olhoeft_g_r_mendell_w_physical_1991}. 
This layer of loose to overconsolidated, fragmented material formed through billions of years of meteoroid bombardment, impact gardening, and thermal breakdown. It consists of fine dust, glassy particles, agglutinates, lithic fragments, and other impact-generated debris \citep{heiken_lunar_1991}.  Previous missions like Apollo, Luna, Lunokhod, and Chang’e sampled regolith from depths ranging from several cm to 3 m using scoops, core tubes, and cone penetrometers  \citep{Heiken:1991, carrier_d_w_olhoeft_g_r_mendell_w_physical_1991}.
 Efforts were made to map the regolith at greater depth, for example, using ground-penetrating radar (GPR) by the Chang’e-3 (e.g., \cite{fa_regolith_2015}) and Chang’e-4 rovers (e.g., \cite{ding_moon-based_2023}). While these instruments were capable of acquiring high-resolution subsurface profiles, the high-frequency channels experienced signal artifacts and scattering effects that reduced interpretability, particularly with respect to fine-scale layering \citep{li_moons_2020, pettinelli_overview_2022}. 
As a result, the lateral and vertical variability of the regolith remains an open question.

Equally uncertain are the regolith’s geomechanical properties, along with their vertical and lateral variability. Parameters such as density, cohesion, internal friction angle, and shear strength remain poorly constrained yet critically influence the load-bearing capacity of the ground, the feasibility of excavation or construction activities, and robotic mobility/energy consumption, often summarized as 'Trafficability'. Trafficability-related properties are particularly important near and along pit rims and funnels, where slope stability governs both scientific access and operational safety. Since the Surveyor and Apollo programs, orbital images were used \citep{carrier_d_w_olhoeft_g_r_mendell_w_physical_1991} to constrain the properties of the near-surface regolith in regions way beyond the few robotic and crewed landing sites, including pyroclastic deposits \citep{bickel_analysis_2019}, the circumpolar regions \citep{Bickel_Kring2020}, and even permanently shadowed, volatile-rich regions \citep{Sargeant_bouldertracks_2020}. Yet, site-specific, in situ measurements of the physical properties of regolith are lacking for the Marius Hills region and for most of the lunar surface beyond past landing sites.

\section{Lava Tubes on the Moon}
\subsection{Formation}

\begin{figure}
    \centering
    \includegraphics[width=0.7\linewidth]{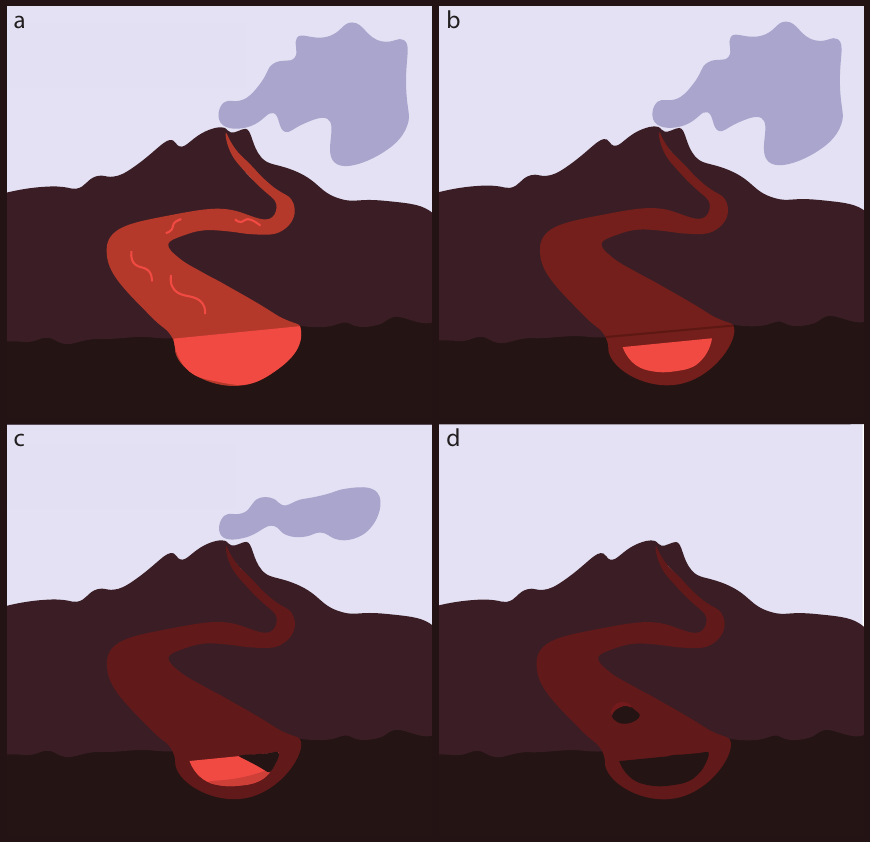}
    \caption{Sketch of a lava tube formation  process where flowing lava becomes thermally insulated beneath a solidifying crust from (a)-(b), until a cavity forms (c), that is finally completely emptied and becomes accessible from the surface via a collapsed ceiling (d). }
    \label{LaveTFormation}
\end{figure}

Lava tubes form when flowing lava becomes thermally insulated beneath a solidifying crust, allowing molten material to continue moving below the surface while the exterior cools (Figure \ref{LaveTFormation}). Two main formation mechanisms are recognized from terrestrial studies: (1) the surface cooling and roof solidification of active lava flows and (2) the inflation and subsequent drainage of a thermally insulated flow, typically occurring farther from the volcanic source \citep{sauro_lava_2020}. Additional pathways include the pressurization of thickening lava flows, deep inflation along older lava flow boundaries, or conduit formation within eruptive fissures. The development and morphology of these tubes depend on factors such as effusion rate, slope, underlying terrain, and lava composition. Once the lava supply wanes, pathways drain, leaving behind arched or elliptical voids. Despite advances in lava-flow simulations \citep{dietterich_benchmarking_2017}, relatively few studies quantitatively address how lava tubes initiate, evolve, and eventually close \citep{keszthelyi_physical_1998, morgavi_tubes_2025}.

The Moon’s lower gravity promotes exceptionally large, stable, and deeply entrenched conduits. Higher effusion rates and the dominance of pressure-driven flood basalt emplacement mean that deep inflation and strong thermal erosion are likely the primary formation mechanisms, producing tubes that could be hundreds of meters wide and tens to hundreds of meters tall. This is one to three orders of magnitude more voluminous than terrestrial tubes \citep{sauro_lava_2020,losarcos_lava_2025}. Their cross-sections tend to be less eccentric due to entrenched morphologies, and collapse chains often occur on very low slopes, sometimes beginning at unusually large vent depressions. 

\subsection{Stability of Lava Tubes and Pits }
\label{stability}
However, the presence of collapse features ('skylights') on the lunar surface suggests that these lava tubes can be unstable under certain conditions.  Many authors have hypothesized formation mechanisms, which include: 1) meteoroid impacts \citep{martellato_numerical_2013}, 2) critical conditions related to the combination of tube geometry, geotechnical properties, and gravitational loading \citep{theinat_lunar_2020,chwala_structural_2024}, and 3) geologic heterogeneity in localized areas leading to anomalously thin roofs, denser than average fracture networks, and/or weaker lithological layers \citep{blair_structural_2017}.  It should be noted that these mechanisms are not mutually exclusive, although studies have used skylight geometry to constrain lava tube size (item 2 above), which assumes continuum failure of the rockmass independent of local heterogeneity.  A more recent study \citep{Williams2025} proposed using subtle topographic bulges running in parallel to volcanic rilles to constrain tube width. These hypotheses have been explored using continuum mechanics modeling, and the results depend  on the geometry, geology and geotechnical parameters of the surrounding rockmass and the lava tubes themselves \citep{blair_structural_2017, theinat_lunar_2020,chwala_structural_2024}.   

As very few direct observations of these critical factors have been made, previous work has had to make many simplifying assumptions, and has treated the influence of geometry using sensitivity analyses \citep{theinat_lunar_2020,chwala_structural_2024,Williams2025}.  These models have parameterized rock mass properties using the Geological Strength Index (GSI) approach \citep{hoek_practical_1997}, with values from terrestrial analogues.  However, there have been limited observations of the fracture networks present in lunar basalts, and there is thus a critical unknown related to rock mass strength of these units. %
Furthermore, it is known from terrestrial analogues \citep{okubo_pit_1998,roche_sub-surface_2001,zhao_failure_2022} that stoping, a discontinuum failure mode that can occur at stress levels below overall continuum failure, is a primary formation mechanism of pits in lava tubes.  Although this has been suggested in previous studies \citep[e.g.][]{robinson_confirmation_2012,wagner_lunar_2022}, it has yet to be quantified in numerical models, likely due to a lack of field data regarding fracture networks within the rock masses that host skylights.

\cite{wagner_lunar_2022} have shown that, in order to gain a sufficient field of view of lunar pits required to map stratigraphy and fracture networks, rovers must be able to traverse great pit slopes. Although slope angles vary among pits and across observation locations within a single pit, approaching observation points as close to the rim as possible optimizes the viewing geometry and may even allow direct imaging of the lava cave interior \citep{Margarit2025}. However, observations of numerous granular flows on the Moon indicate that regolith can fail as a dry granular material \citep[e.g.,][]{kokelaar_granular_2017, bickel_global_2022}. Such failures depend on both the local slope geometry and the geotechnical properties of the regolith \citep{hungr_varnes_2014}.
While previous studies have constrained average regolith properties \citep{mckay_properties_1990, bickel_analysis_2019}, site-specific data remain sparse, particularly at pit rims where slopes are steepest and failure risk is highest. Sensors mounted on a legged robot could directly constrain these local regolith properties in situ, reducing uncertainty in slope stability and enabling safer, more informed mission planning.

\section{Investigating Lava Tubes}
Direct in situ measurements to characterize lava tubes and their surroundings, as described above, imply a high level of complexity—and consequently cost—when the mission is designed to physically access and explore the interior of such lava tubes. However, a combined approach using geophysical methods to probe the subsurface from above, together with high-resolution imaging and spectroscopy to characterize the morphology and mineralogy at the pit, can address many of the key scientific questions.
In particular, the existence, lateral extent, geological setting, and structural stability of a candidate pit can be investigated with a small robotic surface mission equipped with complementary geophysical and remote-sensing instruments. This approach enables the assessment of subsurface void geometry, mechanical integrity, and compositional context without requiring direct entry into the lava tube.
In the following section, we describe such an approach in detail, with particular emphasis on the payload carried by the proposed LunarLeaper mission. In the following section, we provide further information on the mission concept.

\subsection{LunarLeaper Subsurface Investigations}

From a geophysical perspective, a lava tube can produce detectable signatures through several distinct physical property contrasts. The instruments carried by LunarLeaper exploit two primary characteristics: (1) the void space represents an absence of mass, producing a density contrast detectable by potential field methods; and (2) the ceiling and floor are interfaces with high impedance contrast for elastic and electromagnetic waves, detectable by wave-based sounding methods. Additional contrasts — notably in remanent magnetization (exploited by magnetic surveys, discussed below) and in electrical resistivity \citep{Haviland2025} — can in principle also be diagnostic, though those have not been included in the current payload.

\subsubsection{Gravity Surveys}
\label{Sec:grav}
\paragraph{Basic Principles}
To start with potential field investigations, the gravity method is a straightforward and powerful geophysical exploration strategy to target subsurface caves, such as lava tubes. On Earth, such a method is commonly used to support locating buried features, such as sinkholes, tunnels and voids associated with mines, quarries and lava tubes \citep{Deroussi:2009}.  

Following Newton's Law of Gravity, the gravitational attraction $\mathbf{g}$ of a mass body with volume $V$ and uniform density $\rho$ is defined as

\begin{gather}\label{eq:gravnewton}
\mathbf{g}(x,y,z) = -G\rho\iiint_V\nabla_{\mathbf{R}}\left( \frac{1}{R}\right) dv, 
\end{gather}

\noindent where $G=6.6743\cdot10^{-11} \mathrm{N\cdot m^2\cdot kg^{-2}}$ is the Gravitational constant and $R$ is the Euclidean distance between a mass point with elemental volume $dv$ inside the target body and the observation point $P$, with coordinates $(x,y,z)$, at which the gravity field is measured. The gradient with respect to $R$ is denoted as $\nabla_{\mathbf{R}}$.
From equation~\ref{eq:gravnewton} above, it can be seen that the key parameters governing $\mathbf{g}$ are the density $\rho$ and the distance $R$ between the source body and the observation point. Because equation~\ref{eq:gravnewton} is linear in $\rho$, it can also be written for gravity changes $\Delta \mathbf{g} = \mathbf{g} - \mathbf{g}_0$ due to a density anomaly $\Delta\rho = \rho - \rho_0$. Gravity anomalies are measured in the non-SI unit $\mathrm{Gal}=10^{-2}~\mathrm{m/s}$. 
In the case of subsurface cavities such as lava tubes, a negative density contrast is expected due to the mass deficit, resulting in a corresponding negative gravity anomaly. Given that for lava tube detection and characterization, mainly the change and not the absolute value of the surface gravity field is relevant, the investigation does not strictly require absolute measurements and hence a relative gravimeter can be used. 

\paragraph{Forward Modeling}

The spatial variation of the surface gravity field due to a subsurface void can be evaluated by measuring $\Delta \mathbf{g}$ on a moving platform while traversing the density anomaly. At the LunarLeaper investigation site, the planned robotic mission will measure $\Delta \mathbf{g}$ along a traverse-like path extending up to 500 meters from the pit; considering a negative density contrast $\Delta \rho$ of the cave with respect to surrounding rocks of 1500 to 2500 $\mathrm{kg/m^{3}}$, the resulting gravity signature of the tube is expected to be strong enough to be detected.

\begin{figure}
    \centering
    \includegraphics[width=\textwidth,keepaspectratio]{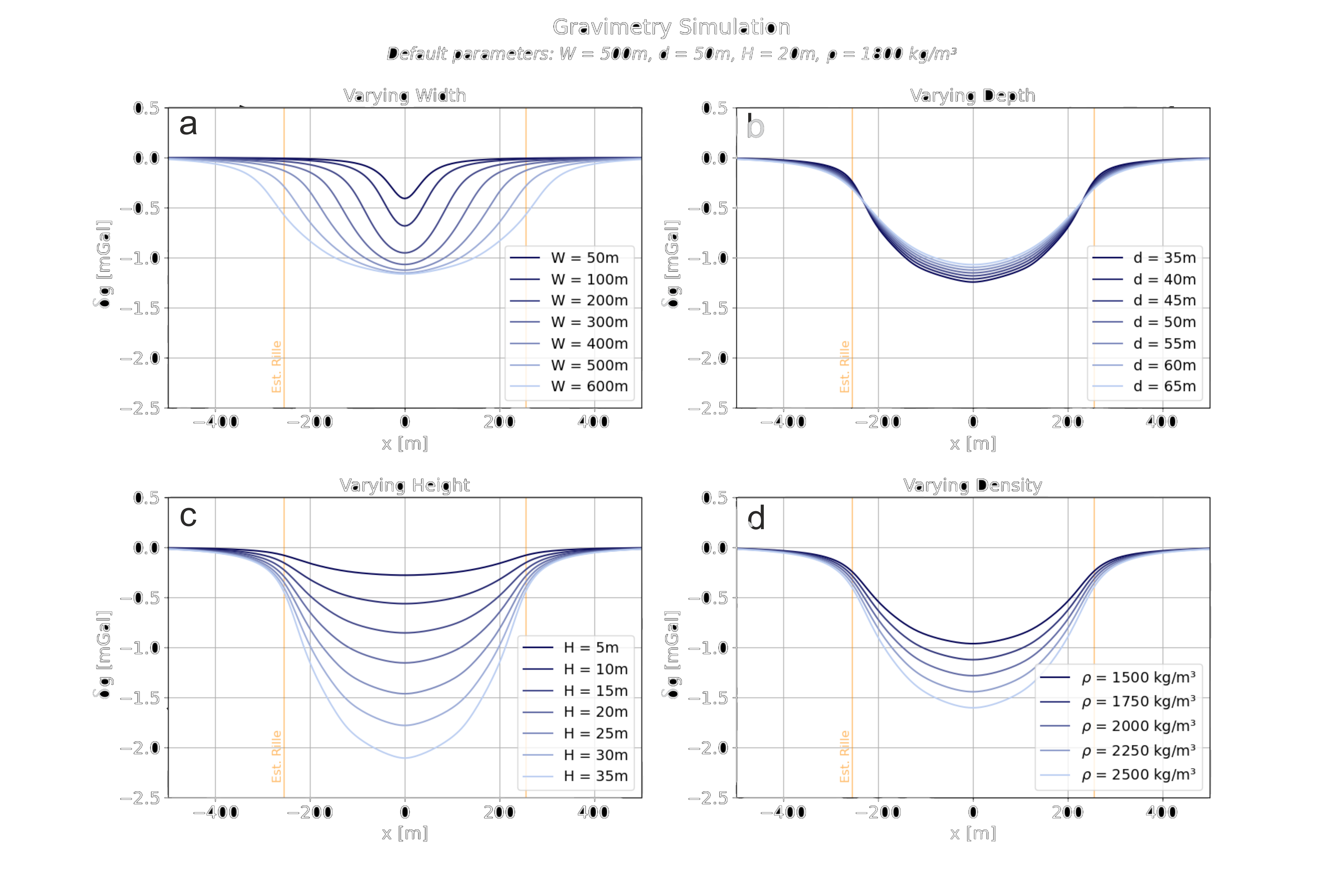}
    \caption{Gravity simulation of a tube with different horizontal diameters ``W'', depths to ceiling ``d'', vertical diameters ``H'', and host rock densities ``$\rho$''. Default values are shown at the top of the figure. We show the gravity responses $\delta g_{z}$ as a function of varying (a) horizontal diameters, (b) depths, (c) vertical diameters and (d) densities.}
    \label{fig:grav_simulation}
\end{figure}

A large lava tube of 500 meters width and 20 meters height located at a depth of 50 meters, and with a density contrast of $\Delta\rho =-1500\,\mathrm{kg/m^{3}}$, would vary the surface gravitation by $\Delta g \approx 1$ mGal compared to the lunar average gravitation of $g_0\approx162.2$ Gal (Figure~\ref{fig:grav_simulation}). Even a smaller tube of just 50 meters width might generate a maximum signal amplitude of $4 \cdot10^{-1}$ mGal when placed at the same depth (Figure~\ref{fig:grav_simulation}). This value is well within the range of the resolution of relative gravimeters used on Earth (e.g., the resolution of the most recent \textit{Scintrex CG-6} is $10^{-4}$ mGal).

\begin{figure}
  \centering
  \includegraphics[width=\textwidth,keepaspectratio]{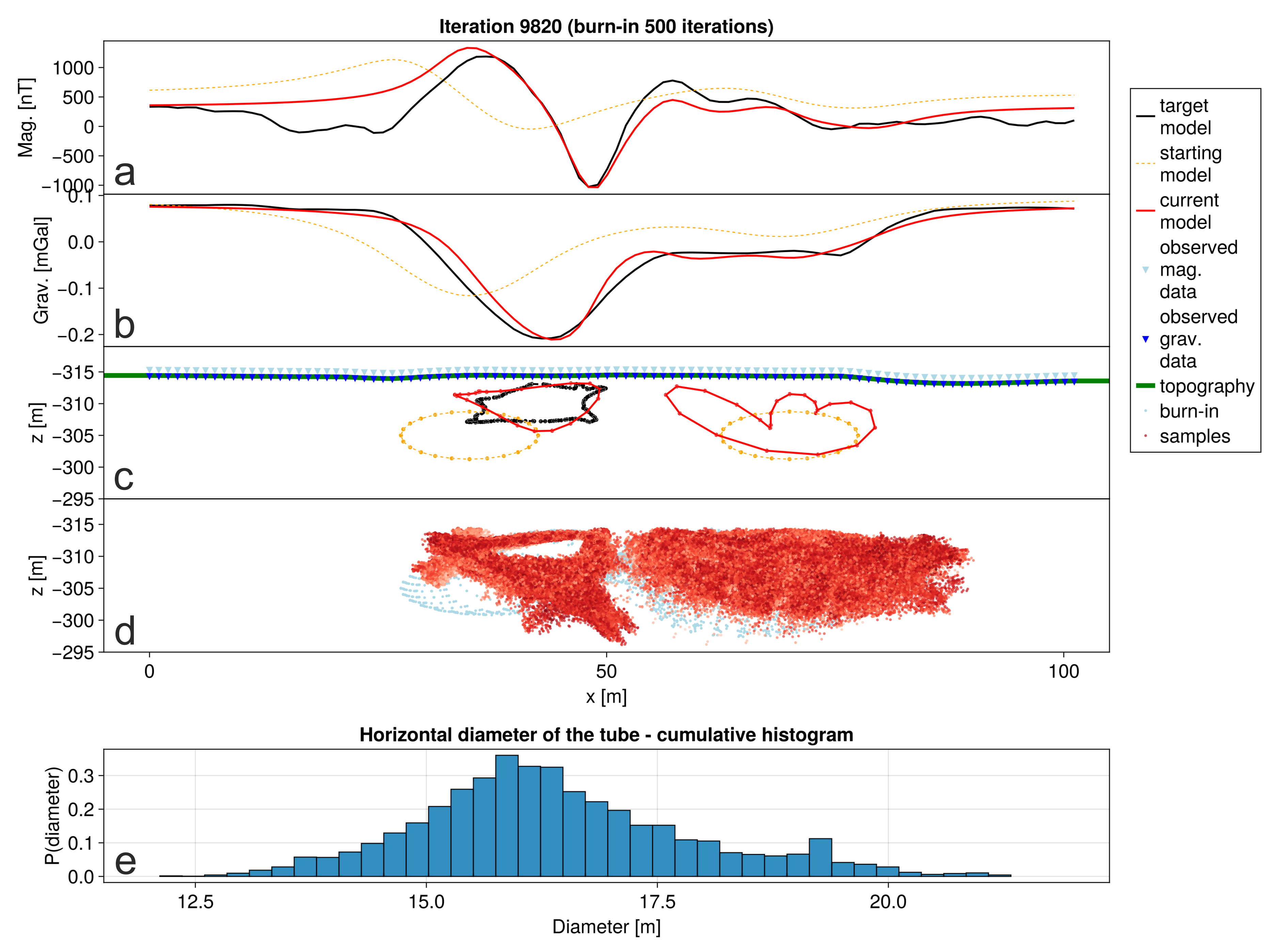}
  \caption{Selected model and data at iteration 9820 of 10000 from the 2D joint inversion of gravity and magnetic anomaly data collected in the ``Cueva de Los Naturalistas'', Lanzarote~\citep{ghirotto2026seismic}. (a) Magnetic anomaly data (black),  calculated responses from the model at the selected iteration (red) and the starting mode (orange), respectively. (b) The same as a) but for gravity data.
  (c) Polygonal bodies: the polygonal cross-section traced along the 2D data profile of the lava tube serving as reference model (black), obtained via a Laser Scanner; note that this is the only cave and the additional small anomalies on the right side of the panels (a)-(b) are from local higher porosity and lower magnetization of basaltic rocks. The starting models are represented by the central prominent lava tube and the aforementioned stratigraphic heterogeneity (orange); the model polygonal bodies at the selected iteration (red). (d) Position of the polygons across the iterations during (cyan) and after (red) the burn-in phase (i.e., initial 500 iterations). (e) Probability of each horizontal diameter explored during the inversion. The highest probable values range between 15 and 17 m and the actual value measured is $\sim$15 m.}
  \label{fig:HMC_gravmag}
\end{figure}

\paragraph{Analogue Studies}
A valuable approach to testing and validating the applicability of the gravity method for investigations on the Moon and other planetary bodies is provided by Earth-analogue structures~\citep[e.g., ][]{montanez_munoz_gravity_2025}. In recent years, Lanzarote island (Canary Islands, Spain) has attracted significant attention from planetary geophysicists, as it hosts well-preserved lava tube systems~\citep{Tomasi2022} embedded in basaltic rocks with composition similar to lunar and martian basalts~\citep{sauro_lava_2020}. 
Very recently, a gravity investigation was performed in the ``Cueva de Los Naturalistas'' in the UNESCO Global Geopark of Lanzarote, a lava tube system located in the central portion of the island. This investigation was part of a multi-method geophysical campaign including passive and active seismic, magnetic and gravity surveys~\citep{ghirotto2026seismic}. At the same time, a Laser Scanner survey was performed inside the tube to retrieve its 3D geometry. 
Analysis of gravity anomaly data showed a prominent negative anomaly above the explored buried lava tube, in agreement with a similar signal pattern in the magnetic anomaly data and the anomalous behavior of seismic waves. Subsequent 2D joint inverse modeling of gravity and magnetic anomaly data, using the probabilistic Hamiltonian Monte Carlo strategy~\citep{zunino2022hamiltonian}, clearly imaged the section of the lava tube along the data profile and provided a precise estimation of its horizontal diameter (Figure~\ref{fig:HMC_gravmag}; methodological details are provided in \cite{ghirotto2026seismic}). These results demonstrate the effectiveness of both the inversion strategy and potential field data for characterizing the geometry and physical properties of subsurface lava tube systems. A discussion focusing on magnetic surveys as also shown in Figure \ref{fig:HMC_gravmag} is provided in sub-Section~\ref{sec:magsurv}. 

While detectability of a tube with a relative gravimeter is not a concern from a physical/mathematical viewpoint on Earth or on the Moon, gravity modeling is inherently affected by non-uniqueness, i.e., a given measured dataset can be fit by the gravity response of an infinite number of models. In particular, gravity observations alone have limited resolving power with respect to the exact geometry and depth of density contrasts in the subsurface. In this context, complementary geophysical techniques that directly constrain geometry of a lava tube, such as the Laser Scanning and Ground Penetrating Radar (GPR; see sub-Section~\ref{sec:gpr}) are particularly useful to reducing model ambiguity and to ensuring geological plausibility of gravity-based interpretations.

\paragraph{Operational Challenges}
A major operational challenge affecting gravity measurements is the sensitivity of relative gravimeters to (i) thermal changes and (ii) instrument tilt. Temperature fluctuations severely affect instrument sensitivity as temperature gradients inside the system can lead to shifting baselines. An instrument tilted by a small angle $\alpha$ will measure part of the (dominant) vertical component $g_z$ of the planetary gravity field $\mathbf{g}$ on its horizontal components (proportional to $g_z\sin \alpha \approx g_z\alpha$).
In contrast, the effect of tilt on the vertical component manifests as a second-order term proportional to
$g_z|\mathbf{g}| \cos \alpha \approx g_z (1-\alpha^2)$, and is therefore orders of magnitude smaller than on the horizontal components. 
On Earth, relative gravimeters typically only measure $g_z$ and tilt is carefully, and typically manually, minimized by leveling the instrument in the field before acquiring measurements at each acquisition point. This leveling can be done with a bubble tiltmeter or by aligning the instrument to a geodetic reference frame. 
For a planetary mission, such leveling procedures are generally not feasible, particularly on moving platforms where leveling would need to be repeated at each measurement location. In these cases, the approach is to measure the full 3D gravity vector and reconstruct the orthogonal components during post-processing.

 The combined impact of these effects is illustrated by the Apollo 17 Traverse Gravimeter Experiment, which, despite a theoretical resolution of 0.035 mGal, achieved an operational performance on the lunar surface of only $\sim$2 mGal \citep{Buck:1973} due to thermal drift and leveling uncertainties \citep{Talwani:1976, center_us_apollo_1972, Urbancic:2017}. Advances in gravimeter technology since Apollo, in particular improved thermal management and leveling-free designs, suggest, however, that future deployments could achieve significantly better performance.

The \textit{GRAvimeter for small Solar System bodies} (GRASS) is an instrument  specifically designed to address these challenges \citep{noeker2022grass}. It was developed in the context of the European Space Agency's Hera mission \citep{Michel:2022} launched in 2024 and is currently on its way to the Didymos binary asteroid system. GRASS is accommodated on Hera's CubeSat \textit{Juventas} which will be deployed in the asteroid system in 2027 and land on the system's moonlet to enable the gravimeter to perform surface acceleration measurements for mass, interior structure, and system dynamics investigations. As the post-landing attitude of the CubeSat is not controlled, GRASS is designed to reconstruct the full 3D vector that directly provides the surface accelerations with an accuracy on the order of a few tens of microgals. The instrument is a miniaturized cantilever-based gravimeter featuring two orthogonally aligned axes and a dynamic measurement. It can be accommodated within a 1U CubeSat form factor. The two cantilever beams are rotated along their long axis to modulate the gravity signal seen by the sensor head, effectively removing thermal drifts and biases \citep{noeker_design_2025}. 
By combining the capacitive measurements of beam deflection amplitudes with the maximum deflection direction along their respective sensor head rotation, the full 3-dimensional surface acceleration vector can be reconstructed. This makes traditional leveling procedures unnecessary.

Beyond instrument sensitivity, a microgravitational survey analysis requires correcting for the contributions of measurement elevation and local topography. Relative elevation changes are corrected by free-air and Bouguer corrections, which can be done employing a shape model of the Moon and assuming a mean density, while for terrain corrections (e.g., effects of surrounding rocks and boulders) a detailed digital elevation model (DEM) of the region is needed \citep{HENRIKSEN2017}. The lunar free air gradient is $\left|\frac{\partial g}{\partial z}\right| \approx 0.187 \ \mathrm{mGal\,m^{-1}}$; hence, to reduce the error in free-air correction to less than 0.05 mGal, a knowledge of the elevation by less than 0.25 m is needed, which needs to be accomplished by a combination of a high resolution DEM and simultaneous location and mapping (SLAM) of the robot.

\subsubsection{Ground Penetrating Radar (GPR)}\label{sec:gpr}
\begin{figure}
    \centering
    \includegraphics[width=\textwidth,keepaspectratio]{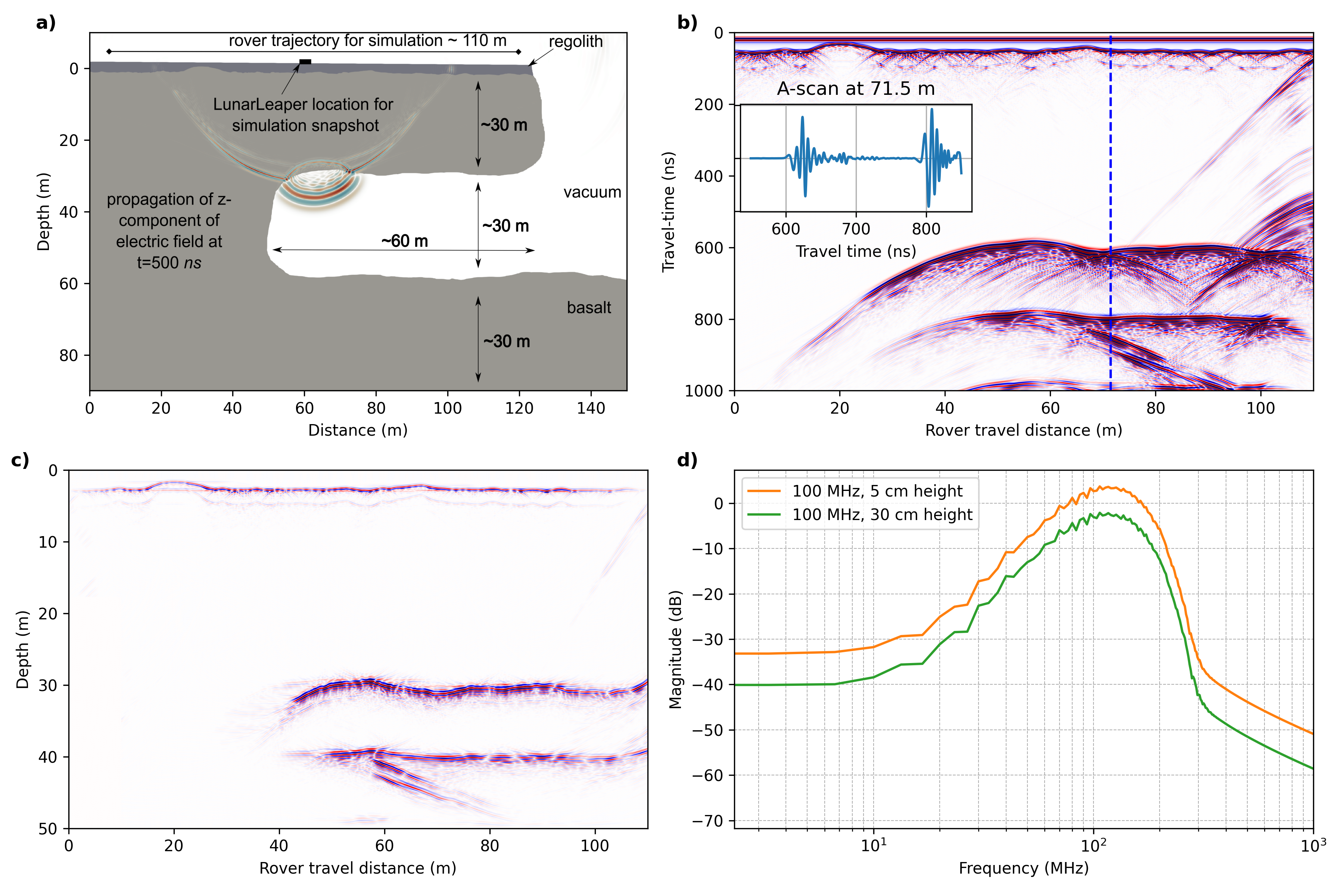}
    \caption{Simulating the electromagnetic wave propagation for the Lunar Leaper antenna. a) The conceptual model used to run the simulations, along with a snapshot of the electromagnetic field with the rover at roughly the middle of its trajectory. b) The "raw" simulated data. Here, we only applied a linear time gain with an exponent of 2 to make late arrivals visible. The insert shows an A-scan along the rover's path. c) The post-processed and migrated data. d) A power comparison between the two simulations, showcasing the increased signal obtained when the antenna is closer to the ground. Note that the rover's trajectory for the simulation, and hence the GPR profiles, ends at the brink of the pit (where the "regolith" arrow points).}
    \label{fig:gpr_simulation}
\end{figure}

\paragraph{Operating Principles}
GPR is a geophysical imaging tool that generates high-frequency electromagnetic waves, typically in the MHz and GHz range, that allow for high-resolution imaging of subsurface structures. In a sounding survey, GPR waves are generated by a transmitter antenna, propagate into the subsurface, and reflect/scatter from dielectric contrasts. The receiver measures the reflected response, and as the whole system moves along a trajectory, an image (profile) is created (Figure \ref{fig:gpr_simulation}). The propagation and scattering of GPR waves depend on the dielectric properties of the medium, namely the dielectric permittivity $\epsilon'$, which controls polarization and therefore wave velocity, and the electrical conductivity $\sigma$, which defines the attenuation that a wave undergoes. The dry and cold environment of the Moon offers particularly good conditions for GPR sounding, as the absence of moisture leads to a low $\sigma$, often several (>2) orders of magnitude greater than on Earth. This results in minimal signal attenuation and hence to deeper penetration of the generated waves. In addition, the interior of a lava tube is a vacuum, which has a dielectric permittivity of 1 and electrical conductivity of zero, providing a large contrast to the surrounding basaltic rock or regolith. As a result, the presence of void spaces, such as lava tubes or buried cavities, produces strong contrasts in dielectric properties relative to surrounding materials, creating especially strong GPR reflections, which are prominent in the resulting profiles \citep{esmaeili_resolution_2020}.

In lunar exploration, GPR has played a prominent role over the years, offering direct insight into the subsurface structure of the Moon. Pioneering efforts such as the \textit{Apollo Lunar Sounder Experiment} (ALSE) conducted from orbit during Apollo 17 demonstrated the feasibility of subsurface probing, revealing features from several hundred meters to a few kilometers \citep{porcello_apollo_1974}. More recent advancements have come from China's Chang’e-3 and Chang’e-4 missions, which deployed rover-mounted GPR systems on the surface to collect high-resolution data from the shallow subsurface down to tens of meters \citep[e.g.,][]{ding_moon-based_2023, fa_regolith_2015}. These datasets have revealed a complex, multi-layered stratigraphy of volcanic deposits and impact ejecta. Notably, Chang’e-4 detected a buried basaltic lens at 27 m depth, possibly linked to the Moon’s most recent volcanic activity (2.5–2.2 Ga). 

\paragraph{The LunarLeaper GPR}
For LunarLeaper, we plan to use a GPR instrument based on the \textit{Radar Imager for Mars’ Subsurface Experiment} (RIMFAX) \citep{Hamran:2020}, the first GPR on Mars mounted on the Perseverance rover.  Operating between 150 and 1200 MHz, RIMFAX has been providing high-resolution imaging of subsurface layers to depths of more than 15 meters on the crater floor and more than 30 meters on the Jezero delta. Aiming towards detection of lunar lava caves expected to be 10s of meters deep, we foresee the lunar RIMFAX in combination with an antenna with significant gain around $\sim$100 MHz. The choice of antenna size and type influences the frequency band, where larger antennas increase expected penetration depth, but also potentially restrict the mobility of the robot. Further, we baseline 10 cm sampling which allows mapping of more surficial structure in addition to cave detection as indicated by current modeling efforts (Figure \ref{fig:gpr_simulation}).

\paragraph{Forward Simulations}
To evaluate the feasibility of a GPR antenna in detecting Lunar caves, we performed forward simulations of wavefield propagation and scattering using a widely used Finite-Difference Time-Domain (FDTD) simulator called GPRMax \citep{warren_gprmax_2016}. We assign a top regolith layer of approximately 3 m, drawn with some expected roughness, and a basalt-like layer of approximately 30 m thickness, above the cave, which itself has a height of 30 m and a width of 60 m (Figure \ref{fig:gpr_simulation}a). Details of the simulation and processing can be found in \ref{sec:appendix_GPR}. The simulation uses subsurface materials representative for the MHP location, as obtained from the Clementine mission's UV camera \citep{lucey_lunar_2000, speyerer_precise_2023}. Our objective with these simulations is threefold: 1) to assess whether the height of the antenna above the ground is expected to play a major role in the data signal-to-noise ratio, 2) to investigate the suitable spacing between adjacent measurements along the path of the rover, and 3) to evaluate which structural aspects of the cave can be expected to appear in the profile.

The result of the GPR simulation is shown in Figure \ref{fig:gpr_simulation}.%
Without any scattering and considerable attenuation, it is evident that the details of the regolith and clear reflections from the top and bottom of the cave arise. The 45$^{\circ}$ line that originates in the upper right of the image is the reflection from the edge of the pit. In the figure insert, we show a single trace (A-scan) obtained at 71.5 m along the rover's trajectory. The trace (insert in Figure \ref{fig:gpr_simulation}b) indicates a phase shift from the bottom of the cave (780 $\mathrm{ns}$) reflection compared to that from the top of the cave (600 $\mathrm{ns}$). This phase shift arises because of a flip in the reflection coefficient, indicating a jump in dielectric properties from the basaltic rock to the vacuum. Such a phase shift may be an indicator that the reflection arises from a cavity.

 We present post-processed and migrated data to facilitate visualization (Figure \ref{fig:gpr_simulation}c), using a Kirchhoff migration with a constant velocity of 0.1 $\frac{\mathrm{m}}{\mathrm{ns}}$, which represents the velocity in basalt with the chosen dielectric properties. After migration we see two prominent reflections at 30 m and 40 m depth. The 30 m reflection is the surface of the cave, and appears where expected. The 40 m reflection is the bottom of the cave and is mislocated at 40 instead of 60 m. The reason for this is that we use a constant velocity for the migration which is 1/3 of the speed of light in vacuum. After being transmitted from the cave's top, the waves travel in vacuum at 3x the speed, and therefore the distance is underestimated by that factor as well, resulting in a compressed cave. In the lack of any clear diffractions from the radargram, it will be challenging to obtain a velocity model for the EM wave propagation. A more complex velocity model for the migration is not in the scope of this study, as this would also not be available in a realistic data processing scheme. However, this is an important point to keep in mind when interpreting the cave's structure from GPR profiles and highlights how information, e.g., from gravity, would be an important secondary constraint.

Further, we compare the expected amplitudes of reflected signals for two antenna heights (Figure \ref{fig:gpr_simulation}d, processing, see sect. \ref{sec:appendix_GPR_proc}). The curves show, as expected, a higher energy reaching the cave when the source-receiver pair touches the regolith floor. We performed further computations by subsampling the original image along the rover's trajectory to simulate a sparse acquisition of A-scans, every 50 cm instead of 10 cm, and while the image resolution was reduced, this affected mostly shallower structure while the cave was still readily visible in the profile.

\subsection{LunarLeaper Surface Investigations}
The geological formation process of the tube additionally leaves clues that are directly visible on the surface or in the wall of a pit, if it is present. Both can be detected by imaging or mineralogical surveys.  

\subsubsection{Camera}

An imaging instrument provides invaluable scientific and contextual information about the morphology and geology of lunar pits as well as their environment. For example, an imager enables the extraction of stratigraphic and structural information from the pit wall and its comparison with the geophysical data acquired by other instruments around the pit. An imager is also able to aid in other surface- and regolith-related investigations while enriching outreach-related efforts.

\begin{figure}
    \centering
    \includegraphics[width=0.5\linewidth]{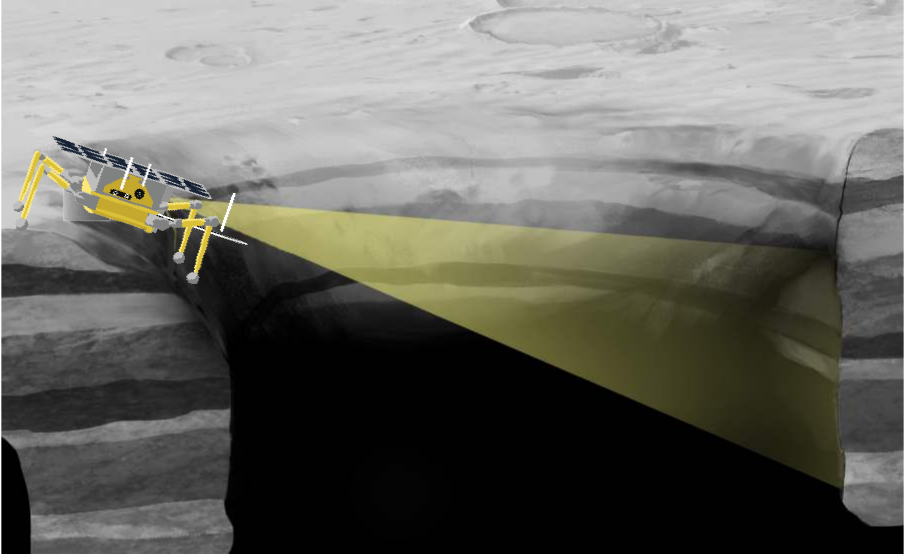}
    \caption{Graphic impression of LunarLeaper observing the near-vertical stratigraphic column of a pit. Not to scale. }
    \label{Fig:Camera_layers}
\end{figure}

At the pits, the LunarLeaper imager will target the stratigraphic column exposed along the near-vertical pit walls. The imager must be able to spatially resolve key elements of the stratigraphic column (Figure \ref{Fig:Camera_layers}), specifically the individual volcanic deposits/flow events, the contacts between them and any potentially present paleo-regolith deposits (Section \ref{volc_evo}). In addition, the imager needs to be able to identify small-scale structural features, such as fractures and faults (Section \ref{stability}). The geometry of the pit and its funnel implies that any observations of the MHP pit wall will require LunarLeaper to image from the opposite side of the pit and funnel, over an extended distance (up to $\sim$100 m). An appropriate field of view (FoV) will help reduce the total number of images required to cover wide swaths of the pit wall, but it does not need to exceed $\sim$45° due to the curvature of the wall. Full-color capabilities would be helpful to distinguish individual volcanic and paleo-regolith deposits. The deeper strata the imager can resolve, the further back in time it would be able to look; ideally, the imager would be able to directly image the cave/entrance that is expected under the MHP, at a depth of approximately $\sim$30 m, without moving on excessively steep (>15°) sections of the funnel surrounding the pit. Along the traverse, the imager will acquire contextual images of the surrounding terrain, providing both geological context and an assessment of local surface conditions.

To resolve structures of $\sim$5 cm in size across the pit from a distance of 100 meters, an angular resolution of 0.25~mrad is needed. This resolution is required to be able to identify fracture networks, allowing to assess the cave stability and layering. This can be accomplished with a commercially available RGB imager with 2048x2048 pixels and $40^{\circ}$ horizontal Field of View, such as the MCAM used by NASA’s Volatiles Investigating Polar Exploration Rover (VIPER) rover as NavCam \citep{rezich_investigating_2025}. The same system is also used as an engineering camera on several European Space Agency (ESA) missions in Earth orbit or deep space (e.g. on BepiColombo; \cite{galluzzi_bepicolombo_2022}).

 With a $40^{\circ}$ FoV, the imager would be able to see parts of the opposite pit wall from an imaging position on the funnel. Preliminary analysis of high-resolution MHP topography suggests that integrating the imager at a height of $\sim$0.7 m above the surface would enable LunarLeaper to observe the top 15 meter of the pit wall without navigating funnel slopes steeper than 5° \citep{Margarit2025}. The same study suggests that LunarLeaper would be able to directly observe the anticipated cave entrance at $\sim$30 m depth if the robot can navigate funnel slopes up to $\sim$10° \citep{Margarit2025}. 
 
 Increasing this resolution could further facilitate the identification of finer stratigraphic details, e.g., thin lava flow deposits and paleoregolith layers, and structural features, e.g., fractures at the cm-scale. For context, paleo-regolith layers that formed over as little as $\sim$13.3 million years in between volcanic episodes, are assumed to be about 5 cm in thickness, based on regolith formation rates inferred for the Marius Hills region \citep{neukum_cratering_2001, hartmann_possible_2007}. Higher resolution imaging could hence increase the time resolution of our reconstruction of Marius Hills' volcanic history.

\subsubsection{Fabry Perot Spectrometer}
\paragraph{Mineralogical Investigations}
To determine mineralogical layering of the lava flows exposed at the pit wall, a spectroscopic investigation will be conducted. Previous studies of the Marius Hills complex indicate that variations in olivine abundance within individual flows are consistent with regional-scale observations, both exhibiting a systematic increase in olivine content toward younger basaltic units~\citep{Besse:2011}. The mid- (MWIR) and long- (LWIR) wavelength infrared spectral ranges are particularly suited to investigate olivine content, as the relevant minerals exhibit diagnostic spectral features in that wavelength range. In particular, silicates show the diagnostic Christiansen feature between 7.5 and 9 \textmu m and Reststrahlen-bands and Transparency features are located between 10 and 13 \textmu m \citep{Shirley:2019,Lucy:2021}. If compositional layering is present along the pit walls, we would expect a shift of the Christiansen feature towards longer wavelengths as olivine content increases while moving up the stratigraphic column \citep{Lucy:2021}.   

We acquired Laboratory spectra to quantify the effects of varying olivine content using Fourier Transform Infrared (FTIR) Bruker Verted 90V spectrometers at the DLR Planetary Emissivity Laboratory. Bi-directional reflectance $R(\lambda)$ for endmember minerals were obtained from 0.4 to 20 \textmu m under vacuum conditions where measurements were performed on granular samples with particle sizes between 63 and 125 \textmu m, similar to particle sizes found in the lunar regolith \citep{Shirley:2019}. As we are mainly interested in the LWIR spectral range, we converted bi-directional reflectance to emissivity $\epsilon(\lambda)$ by approximating $\epsilon(\lambda)=1-R(\lambda)$.

\begin{figure}[ht]
  \centering
  \begin{minipage}[t]{0.49\textwidth}
    \centering
    \includegraphics[width=\textwidth]{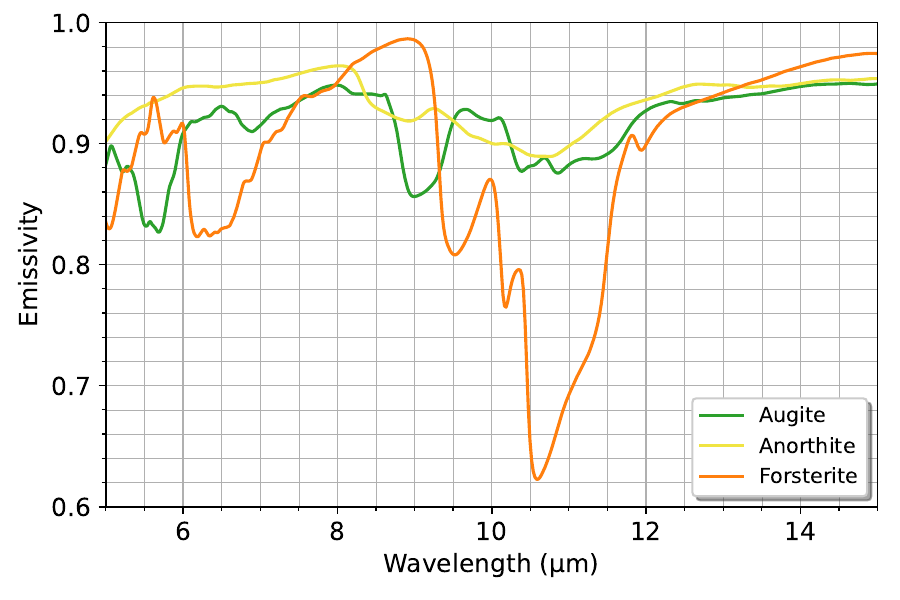}   
  \end{minipage}
  \hfill    
  \begin{minipage}[t]{0.49\textwidth}
    \centering
    \includegraphics[width=\textwidth]{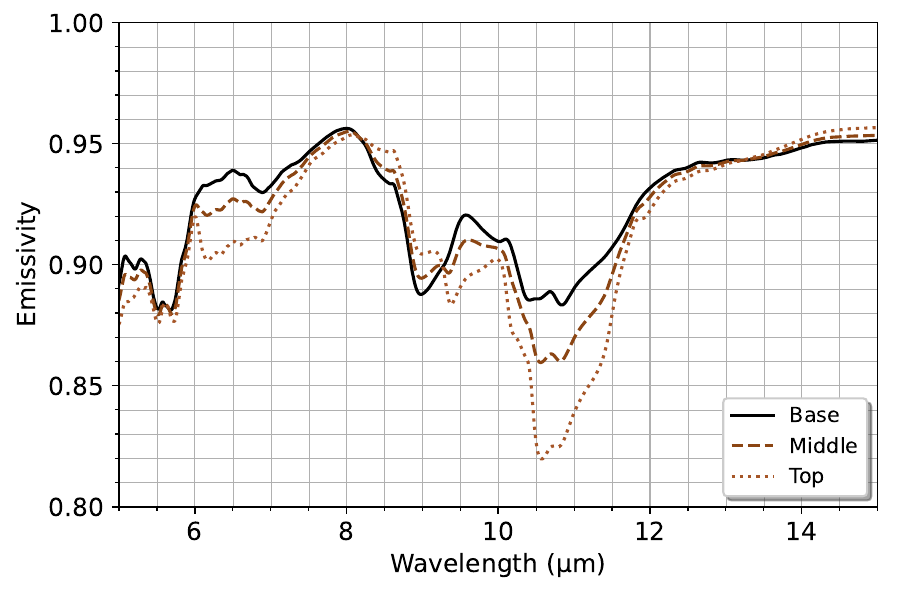}   
  \end{minipage}
  \begin{minipage}[t]{0.49\textwidth}
    \centering
    \includegraphics[width=\textwidth]{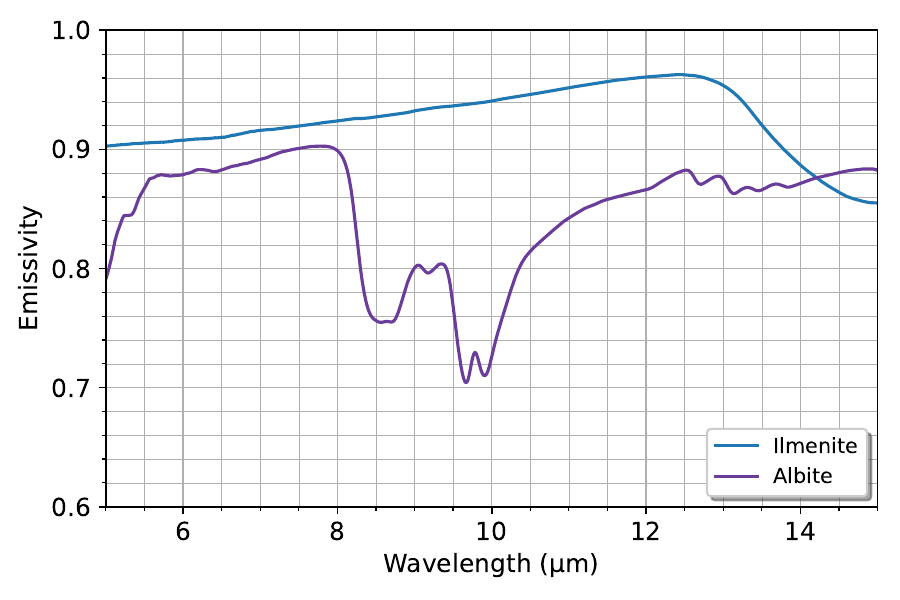}   
  \end{minipage}
  \hfill    
  \begin{minipage}[t]{0.49\textwidth}
    \centering
    \includegraphics[width=\textwidth]{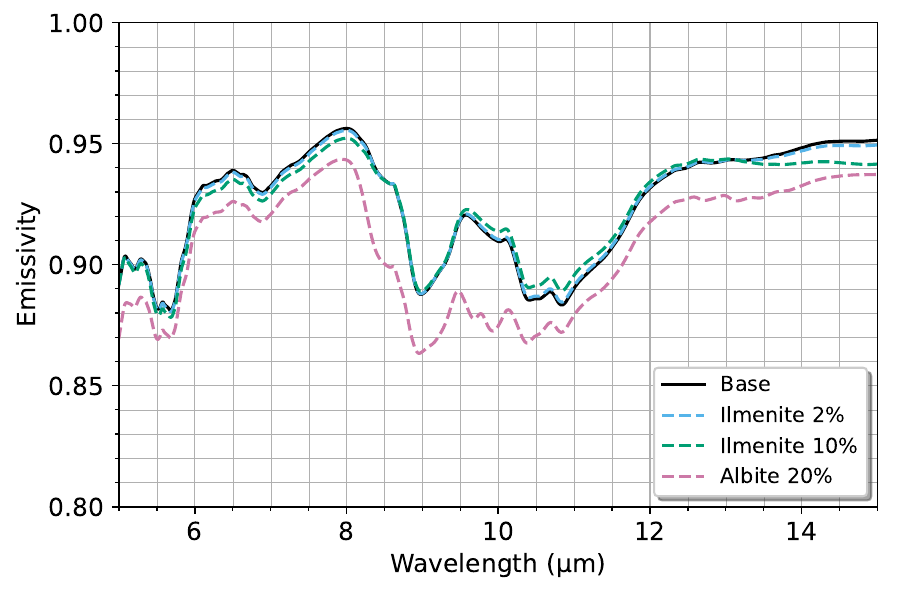}   
  \end{minipage}
  \caption{Emissivity as a function of wavelength for different end-member minerals as well as for mineral mixtures as determined in the laboratory. Top left: End-member minerals used in the mixtures to simulate the top, bottom, and base layers. Top right: Emissivity spectra for mixtures simulating a potential layering in the pit walls with olivine content increasing from the bottom to the top layer.  Bottom left: End-member ISRU materials used in the mixtures. Bottom right: Emissivity spectra for mixtures containing ISRU relevant materials. Measurements were performed under vacuum conditions on granular samples in the 63 to 125 \textmu m size range. See text for mixing ratio details.}
  \label{Fig:LabSpectra}
\end{figure}

As endmember minerals, we assume pyroxenes (augite), plagioclase (anorthite), and olivine (forsterite). To obtain a first order estimate of spectral shapes for mineral mixtures, we performed linear mixing of end-member mineral spectra. 
The mixing ratios were chosen to represent a plausible stratigraphic sequence in lunar mare basalts: 50\% pyroxene and 50\% plagioclase for the base layer, 50\% pyroxene, 40\% plagioclase, and 10\% olivine for the middle layer, and 50\% pyroxene, 25\% plagioclase, and 25\% olivine for the top layer, compatible with ratios found in lunar mare basalts \citep{Heiken:1991}. These mixtures are consistent with a formation scenario in which an initial volcanic plateau formed from silica-rich and olivine-poor magma, followed by olivine-enriched mare flowss, culminating in late mare deposits forming an olivine-rich uppermost layer.

We show results of the measurements on end-member minerals along with mixed spectra corresponding to the base, middle, and top layer in the top row of Figure \ref{Fig:LabSpectra}. For the end-member minerals (top left), the Christiansen feature appears clearly around 8 \textmu m for pyroxenes (augite) and plagioclase (anorthite) and around 9 \textmu m for olivine (forsterite). Also, Reststrahlen-bands are prominent for pyroxenes and olivines between 9 and 12 \textmu m, while transparency features beyond 13 \textmu m are rather subdued, as would be expected at grain sizes around 63-125 \textmu m \citep{Shirley:2019}.

Accordingly, mixtures (top right) show a pronounced Christiansen feature around 8 \textmu m as well as Reststrahlen bands between 9 and 12 \textmu m. Adding olivine and reducing plagioclase when moving up the model stratigraphic column shifts the Christiansen feature from 8 \textmu m (base) to 8.05 (middle) and 8.1 \textmu m (top), while the depth of the Reststrahlen-bands is significantly increased at the same time. Therefore, both spectral regions, i.e., the 7 to 10 \textmu m range as well as the 9 to 12 \textmu m range could serve to identify layering along the pit walls.

In addition to the characterization of layering, spectroscopic investigations could potentially be used to characterize the abundance of materials that are relevant for in situ resource utilization (ISRU).  Relevant materials include albite or anorthite for in situ concrete production \citep{Lee:2022,Zheng:2024} as well as ilmenite for oxygen production \citep{Sargeant:2020,Sargeant:2021}. Starting from a composition of 50\% pyroxene (augite) and 50\% plagioclase corresponding to the base layer model assumed above, we calculated mixed spectra using linear mixing of different fractions of albite and ilmenite to the base layer composition above. Pure mineral spectra are shown in the bottom left panel of Figure \ref{Fig:LabSpectra}, while results of reducing the anorthite content while simultaneously adding 2\% and 10\% of ilmenite as well as 20\% of albite are shown in the bottom right panel.

Due to its rather featureless spectrum in the MWIR and LWIR range, ilmenite has little influence on the mixed spectra. Its Christiansen feature is located around 12 \textmu m \citep{Shirley:2019}, and Reststrahlen-bands are located beyond 12 \textmu m; relatively large amounts of ilmenite would need to be present to be detectable at long wavelengths. For albite, Reststrahlen-bands are pronounced between 8 and 12 \textmu m, and 20\% of albite added to the base layer composition would significantly impact the spectrum between 9 and 11 \textmu m, deepening the respective bands by 2.5\% and enabling identification by spectral measurements.

\paragraph{Instrument}
To fit the mission profile, a miniaturized spectrometer design is essential.  We developed an instrument based on radiometers with flight heritage from numerous planetary exploration missions \citep{Grott:2017,Spohn:2018,Knollenberg:2025}. The spectral capabilities of these radiometers are enhanced using Micro-Electro-Mechanical System (MEMS) based Fabry-Perot filters \citep{Wolter:2025} to generate the necessary spectral resolution. Single pixel thermopile detectors \citep{Kessler:2005} serve as sensing elements, which are particularly suited for the current application due to their broadband absorption and large area, resulting in a large signal to noise ratio. 

The Fabry-Perot filter design is based on bulk micro-mechanical tunable Fabry-Perot interferometers originally developed for infrared gas analysis \citep{Neumann:2008}. The filter's optical cavity consists of two movable mirrors which form the optical resonator, whose optical length can be varied electrostatically. Optical throughput of the filters is greater than 80\% \citep{Ebermann:2016} and filters are characterized by a small size of only 8.5 mm x 8.5 mm. Preliminary shock tests have shown the filter design to be robust up to 1500 g \citep{Meining:2011}. 

Using these filters, spectra are generated in the time domain varying the size of the optical resonator by stepping through electrostatic control voltages between 0 and 70 V while simultaneously measuring radiative flux for each waveband using the thermopile detectors. We operate the interferometers in the first interference order and center transmission around 9 \textmu m with a usable free spectral range between 7.5 and 10.5 \textmu m. The spectral resolution $\lambda/\Delta\lambda$ achieved with the current design is around 30 and noise equivalent emissivity difference is close to 0.5\% \citep{Wolter:2025}. 

\begin{figure}
    \centering
    \includegraphics[width=1\linewidth]{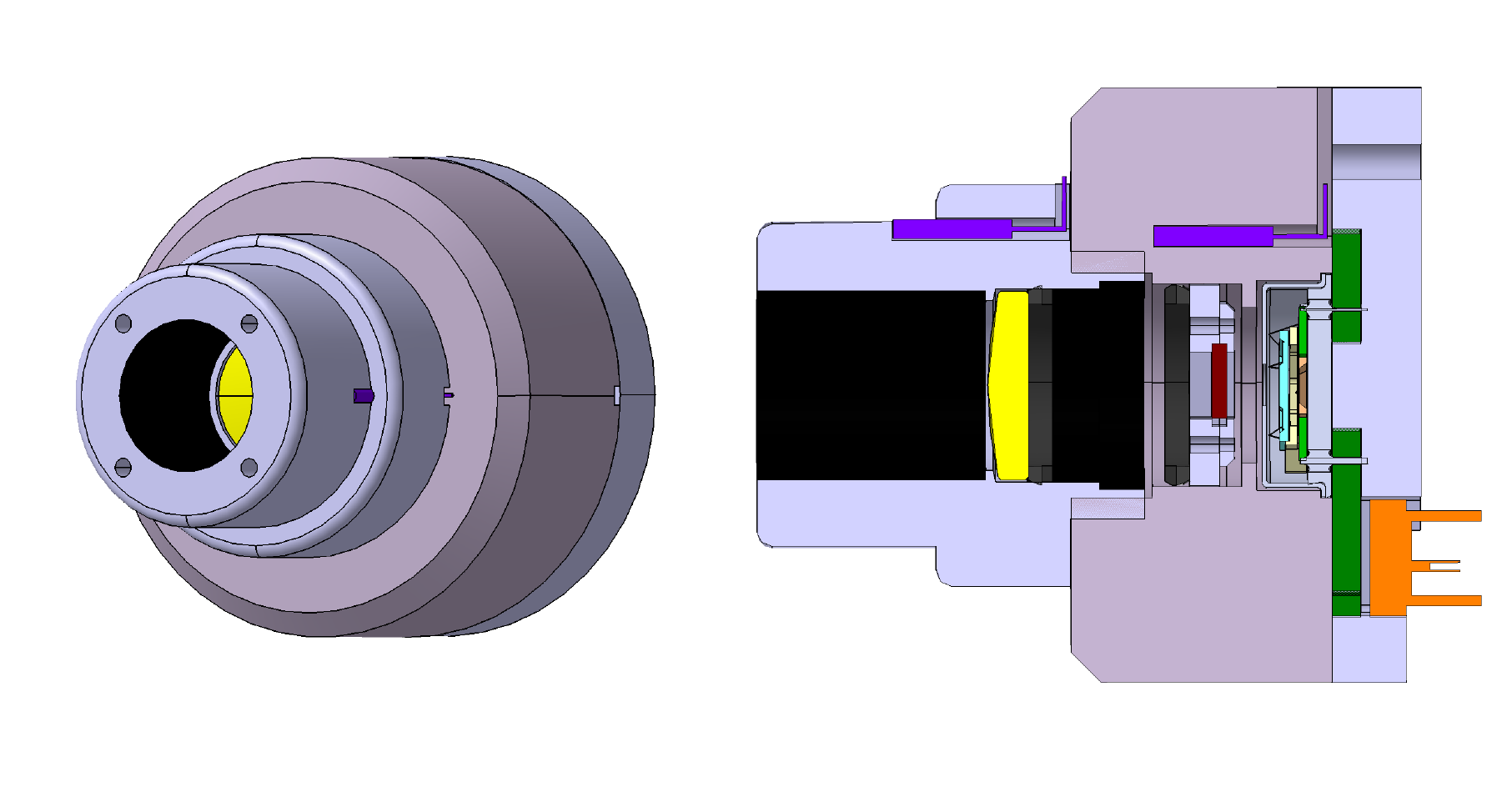}
    \caption{Left: Perspective view of a CAD model of the Fabry-Perot Spectrometer. Right: Cut view of the spectrometer showing (from left to right) the baffle and lens barrel (black), germanium lens (yellow), order selection filter (dark red), Fabry-Perot MEMS interferometer (cyan) and thermopile detector (light brown).}
    \label{Fig:mSpec}
\end{figure}

A computer-aided design (CAD) model of the spectrometer sensor head is shown in Figure \ref{Fig:mSpec}. The sensor head has a diameter of 40 mm, a length of 45 mm, and a total mass of 105 g. Instrument electronics are estimated to weigh close to 200 g and consist of two standard PC104-sized printed circuit boards (PCBs) which accommodate the analogue and digital electronics, respectively. Instrument optics have been designed to allow for a meter-scale ground sampling distance at a distance to target of around 50 m, translating into an instantaneous field of view of 24.5 mrad. The optical tube assembly uses germanium plano-convex anti-reflection coated lenses with a diameter of 12.7 mm and a focal length of 20 mm, which are commercially available (yellow on the left hand side of Figure \ref{Fig:mSpec}). A germanium based order selection (bandpass) filter (dark red) is used to block higher interference orders at short wavelengths and prevents leakage radiation from reaching the detector at longer wavelengths. 

While the current design houses a single filter and detector, the concept can be readily expanded to include two or three filters with a free spectral range of 3 \textmu m each. While adding little mass to the instrument, additional channels would allow for a more detailed characterization of layering as well as increasing the potential for ISRU material identification.

\subsubsection{Geomechanical Experiments}

The articulated legs of the LunarLeaper robot provide an opportunity for dedicated geomechanical experiments that go beyond opportunities offered by wheeled rover systems \citep[e.g.][]{Rezich2025}. Most fundamentally, a leg can be used as a static or dynamic penetrometer, probing the shear strength of the top cms of regolith, although not reaching as deep into the subsurface as the dedicated penetrometers deployed in the 1960s and 70s by the Lunokhod rovers and Apollo program \citep{carrier_d_w_olhoeft_g_r_mendell_w_physical_1991}, as LunarLeaper's feet are designed to limit sinkage. \cite{Kolvenbach2019} demonstrated how legs can also be used to 'probe' the surface in a dynamic manner, directly aiding operational safety. A leg can also be used to dig a trench, providing an opportunity for qualitative observations of regolith cohesion and internal friction angle. A legged robot is also able to manipulate - or pedipulate - its environment, similar to the Apollo astronauts \citep{Arm2024}. On the lunar surface, this capability is useful to flip over small boulders, exposing their more pristine (un-weathered) side. In the vicinity of a pit, pedipulation can be useful, for example, to assess the geomechanical properties of sections of the funnel, such as by tapping on the inclined funnel surface or by moving small pebbles/boulders down the funnel, monitoring their dynamics response.

\subsection{Alternative Geophysical Investigations}
In addition to GPR and gravity, seismic and magnetic techniques offer complementary geophysical approaches to identify subsurface voids. These investigations are not planned for the LunarLeaper mission. 

\paragraph{Seismic Surveys}

Seismically, one straightforward strategy is active reflection imaging, where one uses a seismic source (e.g., a small hammer) and looks for energy reflected from the upper boundary of a lava tube. In principle, the strong impedance contrast between solid rock and void space should generate detectable reflections. In practice, however, the highly heterogeneous uppermost lunar crust scatters seismic waves strongly, reducing coherent reflections and making imaging challenging \citep{blanchette-guertin_investigation_2012}. To overcome this, an array of sensors separated from the seismic source is likely needed to enhance signal‐to‐noise and allow for stacking and migration techniques, but this complicates deployment logistics for robotic or crewed missions.

A second, more deployment-friendly seismic strategy is to remain passive and measure the ambient seismic noise created mainly by naturally occurring thermal moonquakes \citep{civilini_thermal_2023} around 7 Hz frequency. Following \cite{keil_evaluating_2026}, who present a detailed numerical study of different methods, the most promising technique is to analyze the energy ratio of ambient seismic noise on horizontal and vertical components \citep[H/V inversion, see][]{fah_inversion_2003}. This can be done on a single receiver, for example, on a rover or small robot. A substantial void will alter the resonant modes of the subsurface, shifting or introducing resonance frequencies analogous to a subsurface "cavity resonance". By monitoring the ambient noise field and identifying stable spectral peaks and their H/V ratio, it may be possible to infer the presence and approximate depth of a lava tube without needing an active source or a large seismic array. Ideally, this method could be employed during each stop of the rover to detect lateral changes along the traverse. The necessary time for a stable measurement depends on the amount of small moonquakes and the performance of the sensor.
During the NASA InSight mission, this method was successfully used to infer the upper 200 meter subsurface structure \citep{hobiger_shallow_2021, carrasco_empirical_2023}.

If individual thermal moonquakes can be identified in the seismic record and their location estimated, this can be used to obtain surface wave dispersion curves, i.e. the travel speed at different frequencies. If a thermal moonquake occurs within a few hundred meters of the target region, the dispersion curve of Rayleigh waves traveling over a void-bearing structure will differ from those over a homogeneous medium \citep{keil_evaluating_2026}. By inverting the measured dispersion curve, one can infer sharp changes in shear-wave velocity that are consistent with a large cavity or low-density zone at depth. This method requires only a modest aperture between source and receiver but depends on the occurrence and suitable geometry of moonquakes and is hence rather suited for missions longer than just one lunar day.

Passive seismic methods do not require a source component, limiting power consumption. However, seismometers have a certain intrinsic weight, which is inversely proportional to the frequency range. Measurements at a frequency of 1-10 Hz, corresponding to depths of tens of meters, could  plausibly be achieved by MEMS-based sensors with a mass of a few 100 grams. If seismic signals are sufficiently strong, such measurements might even be possible with an onboard IMU \citep{lewis_surface_2019}, although the exact signal levels, such as those associated with thermal moonquakes, remain poorly constrained.

\paragraph{Magnetic Surveys}\label{sec:magsurv}
In addition to gravity measurements, magnetic observations provide a complementary potential-field constraint.  
In an analogous manner to gravity (Section \ref{Sec:grav}), magnetic measurements respond to contrasts in remanent magnetization rather than density; however, unlike gravity, magnetic fields decay more rapidly with distance and depend on both the magnitude and orientation of the magnetization of the source body. As a result, combining gravity and magnetic observations has been shown to mitigate challenges associated with non-uniqueness in potential-field interpretations \citep{Mittelholz2025}.

A lava tube represents an absence of material and therefore a local absence of remanent magnetization relative to the surrounding basalt, provided that the host rock is magnetized (see, for example, the analog experiment in Figure \ref{fig:HMC_gravmag}). 
 In contrast to gravity, the applicability of magnetic detection thus depends critically on the surrounding lava having acquired significant magnetization, for example during emplacement in the presence of an active lunar dynamo. 
The Moon has been proposed to generate a long-lived dynamo field \citep[e.g.,][]{Tikoo2017}, although this notion has been questioned more recently \citep{Tarduno2021}. 

For the example of the Marius Hills region, the bedrock is dated to approximately 3.3 Ga \citep{wagner_distribution_2014} and maria are generally older than 2 Ga \citep{Wilhelms1987}. The absence of a magnetic signal associated with the basaltic layer would therefore represent an important constraint on the history of lunar magnetism; however, such a result would not provide information on the presence or absence of a lava tube. 
Consequently, magnetic detection of lava tubes is primarily applicable in regions where basalts are known or expected to be strongly magnetized. Because short-wavelength magnetic features cannot be resolved from orbit, robust predictions of local magnetization based on satellite observations alone remain challenging.

Where applicable, magnetic measurements offer higher sensitivity to small-scale subsurface structures than gravity, owing to the availability of vector information.
Magnetometers are small instruments with low power consumption. However, electric currents onboard the lander and rover generate magnetic perturbations that must be carefully mitigated and, where not possible, characterized prior to the mission so that such signals can be accounted for. This represented a significant effort during the InSight mission \citep{johnson_crustal_2020, mittelholz_origin_2020}.

\section{LunarLeaper - A Mission Overview}

\begin{figure}
    \centering
    \includegraphics[width=0.7\linewidth]{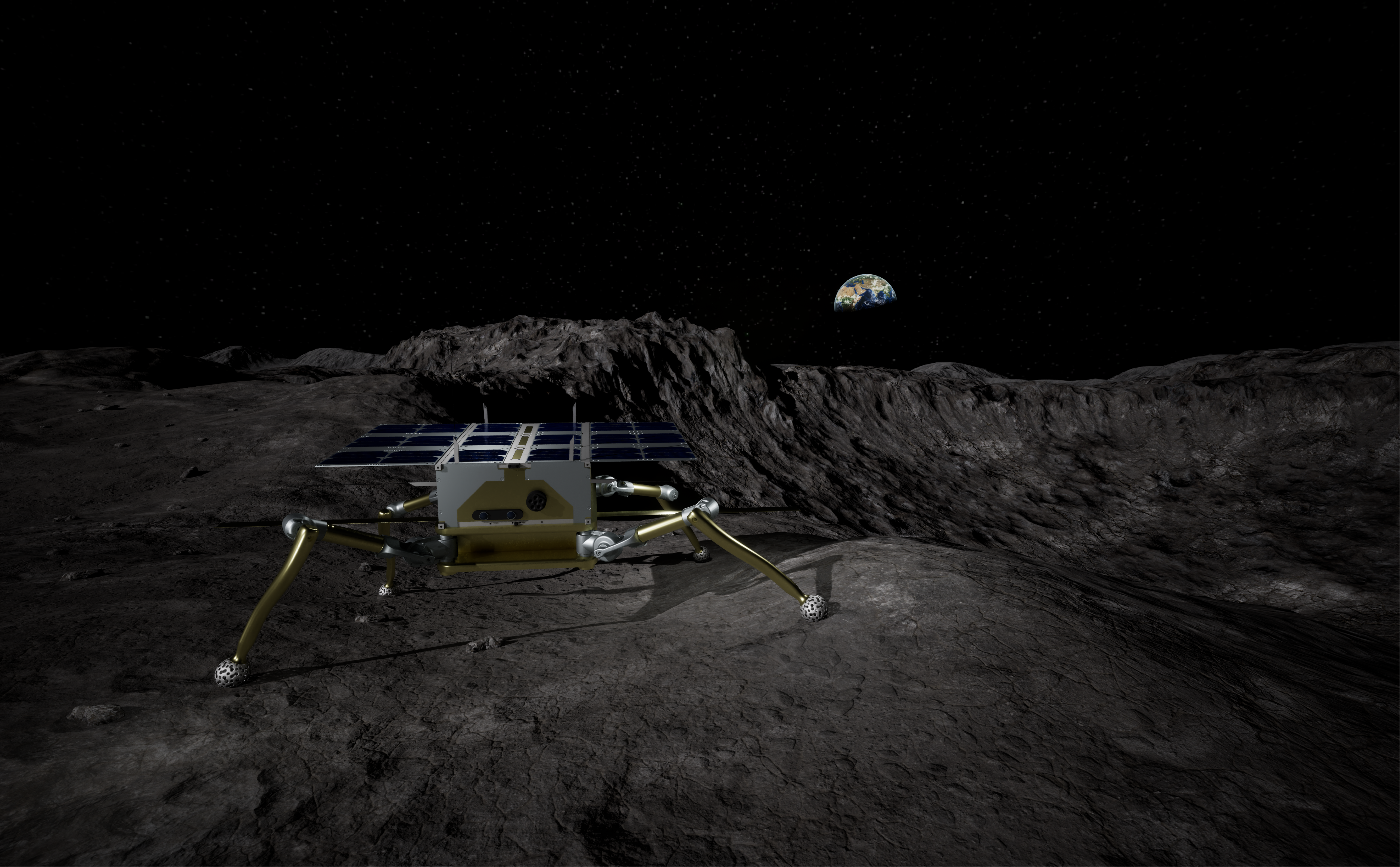}
    \caption{Illustration of the LunarLeaper robot  }
    \label{LL}
\end{figure}
LunarLeaper (Figure \ref{LL}) is a mission concept designed to substantially advance our understanding of the Moon’s subsurface and geologic evolution by exploring mare pits and their hypothesized underlying lava tubes \citep{kolvenbach_lunarleapermission_2026, mittelholz_lunarleaper_2024}. 
LunarLeaper is a legged robot capable of traversing challenging terrain as expected around pits. Unlike traditional wheeled rovers, which struggle with slopes, boulder fields and loose regolith, and the high cost of large, more complex systems \citep[e.g., ][]{miaja_robocrane_2022,hooper_leto_2023}, LunarLeaper’s versatile design enables it to explore pit edges and navigate in challenging terrain autonomously. This capability allows LunarLeaper to investigate key scientific questions about the Moon’s volcanic history and evolution.
While several mare pits offer suitable landing sites for LunarLeaper \citep{Margarit2025}, we focus on the MHP as our preferred landing site in the context of this paper. The MHP is located in a sinuous rille traced by topography that extends 10s of km (Figure \ref{MH_Area}) with a width of approximately 500 m. 

\subsection{Mission Objectives}
LunarLeaper aims to address the following science objectives O1 – O4 (see Science Traceability Matrix in \citep{kolvenbach_lunarleapermission_2026}): 

\textbf{O1: Investigate subsurface lava tubes.} The primary objective of the mission is to confirm the existence and assess the extent of subsurface voids, probing whether lava tubes could be available as a potential site for future investigations. By combining high resolution local ground surveys with available orbital observations, we can gain further insight on regional geology, thus going beyond a local study of the selected landing site / pit. In addition, the results can be used in the analysis of orbital predictions of similar structures elsewhere on the lunar surface.

\textbf{O2: Assess the suitability of lava tubes for human exploration and habitation.} Lava tubes are hypothesized to span hundreds of kilometers, yet only 10 pits have been located in the Mare regions. This raises the question of why the collapse occurred and whether further collapse is likely. A further critical aspect in the assessment of suitability for a lunar base is the availability of in situ resources (ISRs). Lastly, the capabilities of the legged robot allow us to approach the pit and determine the accessibility of the pit edge with potential future large robotic systems or other equipment to be lowered into the pit. 

\textbf{O3: Assess geological processes, with a focus on the volcanic evolution of the Moon.} The exposed stratigraphy along the pit wall provides a unique opportunity to study key aspects of lunar volcanism and interior evolution. By examining mare emplacement processes, LunarLeaper aims to determine the number of flows, their volume, and their timescales, which are central to understanding the Moon's volcanic history.
Basalts layers and changes in composition along the exposed wall offer critical insights into the compositional evolution of the lunar interior and ancient melt sources. Further, the exposed stratigraphy represents mare layering in the overall region and connecting these observations with studies of the geophysically inferred shallow subsurface allow to extend findings from specific sites to more regional context.

\textbf{O4: Investigate the local and regional extent of the regolith.}  Regolith, including paleo-regolith is directly linked to the geological and impact history of the Moon, and its composition and structure holds crucial information. The lateral and vertical extent of regolith is still an open question, one that can ideally be addressed using a mobile robot with a ground-coupled GPR for subsurface investigation. The properties of lunar regolith, including its composition, structure, and mechanical behavior, directly impact mission design, landing safety, mobility, and the feasibility of in situ resource utilization (ISRU) for future lunar exploration as discussed in the context of O2. 

In addition, we define a technical and an outreach objective to highlight the importance of these aspects.

\textbf{O5: Demonstrate the possibilities of dynamic walking robots under lunar conditions.} Legged robots have never been deployed on an extraterrestrial surface, and we intend to showcase a range of aspects that such robots enable including different gaits. 

\textbf{O6: LunarLeaper shall provide 360 degree outreach images along the traverse.} The mission shall be documented to allow for the general public to easily engage with our mission and inspire future generations. 

\subsection{Concept of Operation}
After landing and taking advantage of the currently offered lunar landing infrastructure, we propose to approach the chosen pit with a legged robot, which can access complex terrain and steep slopes that are observed at the pit edge. 

\begin{figure}
    \centering
    \includegraphics[width=0.9\linewidth]{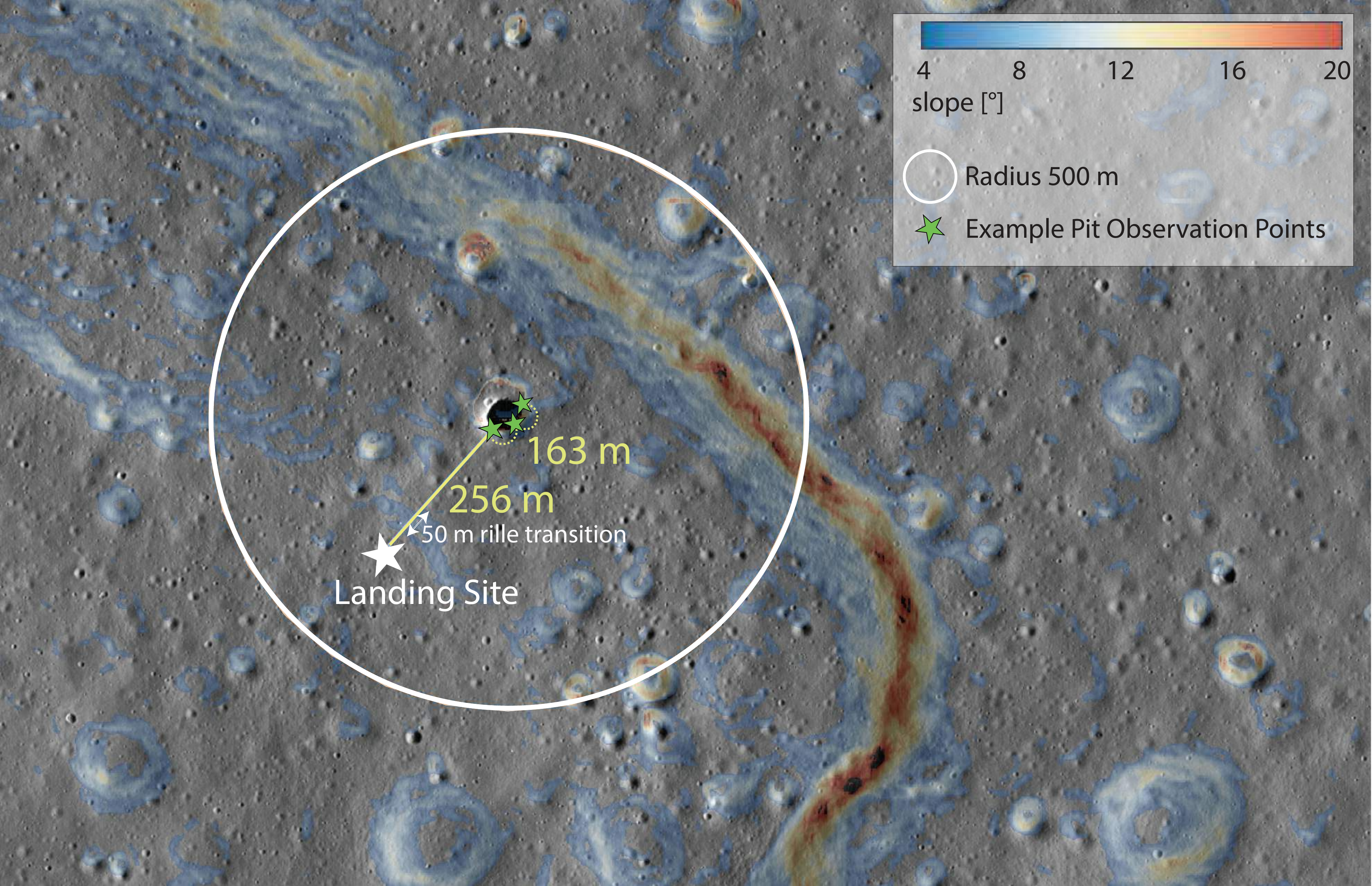}
    \caption{The baseline mission traverse plan plotted on a LRO WAC/NAC mosaic, where slopes between 4 and 20 degrees are highlighted to visually trace the rille. The direct distance between a possible landing site (white star) and the pit is 256 m. Pit observation points (green stars) are exemplary and following the yellow dashed line around the pit adds up to a traverse distance of 163 m.   }
    \label{ConOps}
\end{figure}

LunarLeaper will land on the lunar surface and traverse the rille, possibly the extent of the subsurface lava tube (Figure \ref{ConOps}). 
In this phase, and as expected in the mare region, relatively flat terrain with few obstacles will be encountered while taking geophysical measurements. 
On its traverse the robot will take geophysical measurements that will allow us to survey the subsurface structure and detect and map lava tube geometry if a tube is present. A gravimeter will obtain measurements every 10 m, the GPR will obtain measurements every 10 cm. While traversing the rille which might reflect the edges of the underlying lava tube, the GRAV sampling frequency will be increased to every 5 meters for 50 meters.
In addition, high resolution images facing forward or towards objects of interest along the path will be taken every 20 m. LunarLeaper will autonomously navigate between way points until it reaches the pit. Along the path, cool down stops will be required. This traverse phase concludes once LunarLeaper reaches the pit, i.e., it is less than 10 m away from the pit wall or at a point where the slope is greater than 10 degrees.

After the traverse, the robot will approach the pit edges and acquire high resolution images and compositional information of the pit walls, which contain uniquely exposed layers of the previously geophysically mapped lava flows and regolith layers. It is possible (but not required) that images of the cave can be acquired from the pit edge \citep{Margarit2025}. As stated above, the resolution of the camera is driven by the requirement that the fracture network exposed at the opposite pit wall can be mapped down to the cm level (i.e., at least 10 cm resolution at a distance of 100 m, a margin of 200\% over the diameter of MHP, which is approximately $50\times60$ meters). 
To gain a view of the pit wall, the robot will move to 3 points close to the pit edge (green stars in Figure \ref{ConOps}) and acquire imagery of at least half of the circumferential extent of the pit. 
 At the first vantage point, images of the pits will be taken in addition to a compositional depth profile from the spectrometer; the next two stops serve for a more complete imaging of the pit wall. If time allows, the robot might proceed to map the entire pit, but this would be pursued in an extended mission. 
 This investigation and associated images will allow scientific advances in not only lunar volcanism and regolith formation but also evaluation of the pit structural stability and its use as a possible lunar base. In total, the mission is expected to last one lunar daylight period (14 Earth days with margin).

\subsection{Robot Design}
LunarLeaper is a small robot ($\simeq$30 kg) composed of typical supporting subsystems (power, thermal, structure, avionics, communication) as well as a navigation and a locomotion subsystem. Each subsystem is optimized to maximize operations on the lunar surface.

\textbf{Power \& Thermal:} LunarLeaper relies on solar panels and batteries for power, and a radiator to dissipate the heat. The available power and internal temperature directly dictate the robot's operating duty cycle, or how often and how much scientific measurements can be taken. Design efforts are focused on starting the mission as soon as possible in the lunar day and on reducing the amount of cool-down and recharge periods. The current design allows the robot to reach the pit halfway through the mission, thus guaranteeing several days of operational margin.

\textbf{Avionics \& Communication:} The robot communicates with the lander, which in turn communicates with the Earth. High data rate contact opportunities, to downlink scientific data, are limited, possibly irregular and will have to be shared with the other payloads deployed by the lander. To limit the constraint on scientific measurements, LunarLeaper is equipped with an on-board memory capable of storing all the scientific data generated during the mission. It is expected that 29 GB of (uncompressed) scientific data will be collected in total. This data will be stored on-board and downlinked when contact opportunities arise.

\textbf{Navigation \& Locomotion:} One of the unique aspects of LunarLeaper is that it is a legged robot. While this allows the robot to cross steep and uneven terrain \citep{arm_scientific_2023}, it also acts as an orientating mechanism for all the scientific payloads. Indeed, its 12 actuators allow the robot to pan, tilt or roll in the payload's desired direction. It can also easily rotate on itself, e.g. to take a $360^\circ$ picture. The legged architecture provides significant advantages for mobility compared to wheeled platforms. LunarLeaper is designed to safely operate on slopes up to $\pm30^\circ$, enabling access to rugged volcanic terrains and pit-rim environments that are inaccessible to conventional rovers. Because locomotion relies on discrete foot placement rather than continuous traction, moderate slipping does not prevent forward progress; even if local slip occurs, the robot can recover by adjusting subsequent steps. The platform can also step over rocks and boulders allowing traversal across blocky regolith surfaces. The nominal walking speed is approximately $0.3$~ms$^{-1}$, with peak speeds of up to $0.7$~ms$^{-1}$ depending on the terrain conditions.
Additionally, the robot is trained to walk in such a way that the amount of kicked-up dust is minimized. This is necessary to reduce the amount of dust that will fall down and cover the radiator and solar panels, as well as the instruments' lenses. 

\subsection{Data Product Policies}
All scientific data products generated by LunarLeaper, including gravity measurements, ground-penetrating radar profiles, high-resolution imagery, and spectroscopic observations, will be archived following established planetary science data management standards. Data and associated metadata will be formatted according to community-accepted standards (e.g., Planetary Data System – compatible formats) to ensure interoperability and long-term preservation. After initial validation and calibration, datasets will be publicly released through an open-access archive to ensure timely availability to the wider scientific community.

\section{Outlook}
The next decade offers a unique window for the exploration of lunar pits and lava tubes. Renewed access to the lunar surface through commercial and governmental missions, combined with advances in autonomous mobility and compact geophysical instrumentation make targeted pit exploration increasingly feasible.
Several mission concepts, including the Leto concept \citep{hooper_leto_2023}, RoboCrane \citep{miaja_robocrane_2022} or a previous gravity mapping concept on a rover \citep{carroll2015exploring}, have already highlighted the strong scientific interest in characterizing subsurface voids from the surface. These concepts typically rely on complex architectures and correspondingly greater mission resources.

LunarLeaper intentionally adopts a different approach. As a comparatively small and focused mission, it prioritizes technical feasibility and near-term implementation while still addressing key scientific questions. Rather than attempting comprehensive subsurface access or traversability over large distances with rovers, LunarLeaper targets pit-edge reconnaissance and in situ measurements that can substantially reduce uncertainty about the structure and accessibility of these environments.
In this sense, LunarLeaper represents a pragmatic first step toward the direct investigation of subsurface voids while simultaneously serving as a pathfinder for more capable future platforms. Follow-on missions, potentially employing advanced legged robotic systems, could build on this foundation by extending operations from pit-edge observations to controlled descent and ultimately interior cave traversal, enabling direct study of protected subsurface environments.

in situ exploration of pits and lava tubes has the potential to transform our understanding of lunar volcanic history, crustal structure, and subsurface volatile preservation by providing access to geological exposures and environments unavailable at the surface. Continued development and testing of geophysical techniques in terrestrial analog sites will be essential to reduce risk and maximize scientific return.
Beyond the Moon, technologies and strategies developed for pit and lava tube exploration are directly applicable to Mars and other planetary bodies where subsurface voids may preserve records of geological and climatic evolution. As such, early lunar missions will not only address fundamental lunar science questions but also establish a framework for subsurface exploration across the Solar System.

\section*{Acknowledgments}
AM acknowledges funding through SNF Ambizione grant number PZ00P2\textunderscore209123. A.G. acknowledges the Swiss National Science Foundation (SNF) postdoctoral program (grant \#TMPFP2\textunderscore233881).

\clearpage
\appendix
\section{GPR Simulation Details}
\subsection{Simulation Parameters}
\label{sec:appendix_GPR}

For simulations, we defined a conceptual modeling domain that preserves the dimensions of the expected cave (Figure \ref{fig:gpr_simulation}a). We assign a top regolith layer of approximately 3 m, drawn with some expected roughness, and a basalt-like layer of approximately 30 m thickness before reaching the top of the cave. The cave itself has a height of 30 m and a width of 60 m. The modeling domain was discretized at 5 cm in both vertical and horizontal dimensions, while the third dimension was not modeled, resulting in a 2D simulation. This results in an overestimation of signal amplitudes due to the lack of geometrical spreading in the GPR simulation.

We assign Perfectly Matched Layer (PML) boundary conditions that effectively dampen waves that reach the boundaries of the domain, and simulate an antenna at both 5 cm and 30 cm above the top of the regolith. As a source we assign a ricker wavelet with a central frequency of 100 MHz. We additionally tried simulations with 250 MHz central frequency, and while the spatial resolution increased, the wave attenuation was considerably higher. To ensure detection, we therefore opt for lower frequencies in the LunarLeaper mission.

The source-receiver pair is separated by 1 m for simulation reasons, and moves along the horizontal direction (common-offset) from 10 m to 120 m in the domain (Figure \ref{fig:gpr_simulation}a). A trace (A-scan) is collected every 10 cm. 
For material properties, see table \ref{tab:material_properties}.
The values were back-calculated from a loss tangent of $\mathrm{tan} \theta=0.05$, with $\log_{10}(\mathrm{tan} \theta) = 0.038(\%TiO2 + \%FeO) + 0.312\rho - 3.26$ \citep{heiken_lunar_1991}, assuming an average density of $\rho = 2~ \mathrm{g}/\mathrm{cm^{3}}$ and Titanium and Iron percentages of 4.9\% and 21.4\% respectively, as obtained from the Clementine mission for the MHP location \citep{lucey_lunar_2000, speyerer_precise_2023}.
\begin{table}[b]
    \centering
    \begin{tabular}{l|c|c|c}
        Material & dielectric   & conductivity  & loss tangent  \\ 
        & permittivity $\epsilon'$ &$\sigma \left[\frac{\mathrm{mS}}{\mathrm{m}}\right]$ & 
        \\ \hline
        Vacuum & 1 & 0 & 0 \\
        Regolith & 5 & 0.7  & 0.05 \\
        Basalt & 9 & 1.2 & 0.05 \\
    \end{tabular}
    \caption{Material properties for the GPR simulation}
    \label{tab:material_properties}
\end{table}

\subsection{GPR Processing Steps}
\label{sec:appendix_GPR_proc}
We minimally process the data by applying a horizontal averaging filter over a window of 100 traces, which effectively removes the direct wave visible at the top of Figure \ref{fig:gpr_simulation}b. Following this, we apply a linear gain with an exponent of 2 to compensate for geometric spreading. 

For the amplitude estimate in Figure \ref{fig:gpr_simulation}d, we first take the raw data and scale the entire image so that the maximum amplitude of the electric field is 1 $\mathrm{V}{m}$. We then choose a window of the data for t > 550 $\mathrm{ns}$ to focus on the reflections from the cave and calculate the energy in this window in the frequency domain 

\begin{equation}
P_{\mathrm{dB}} = 20 \log_{10}\!\left( \frac{1}{1100} \sum_{k=1}^{1100} |\hat{A}_k| \right)
\end{equation}
where
\begin{equation}
\hat{A}_k = \mathrm{FFT}(E_{z})_k.
\end{equation}
Here, $E_{z}$ denotes the amplitudes of the electric field component perpendicular to the modeling plane, $k$ is the index to denote an A-scan out of a total of 1100, and $\mathrm{FFT}$ is the Fast-Fourier Transform. 

\bibliographystyle{elsarticle-harv}
\bibliography{LunarLeaperII,references}

@inproceedings{m_s_wagner_r_v_occurrence_2021,
	title = {{OCCURRENCE} {AND} {ORIGIN} {OF} {LUNAR} {PITS}: {OBSERVATIONS} {FROM} {A} {NEW} {CATALOG}.},
	volume = {2548},
	language = {en},
	publisher = {LPI},
	author = {{Wagner, R V} and Robinson, M S},
	year = {2021},
}

@inproceedings{qian_longest_2021,
	title = {{THE} {LONGEST} {SINUOUS} {RILLE} {ON} {THE} {MOON} ({RIMA} {SHARP}) {AND} {ITS} {RELATIONSHIP} {TO} {THE} {YOUNG} {PROCELLARUM} {MARE} {BASALTS}: {IMPLICATIONS} {FOR} {THE} {CHANG}’{E}-5 {RETURNED}},
	language = {en},
	number = {2548},
	author = {Qian, Yuqi and Head, James W and Wilson, Lionel and Xiao, Long},
	year = {2021},
}

@article{neukum_cratering_2001,
	title = {Cratering {Records} in the {Inner} {Solar} {System} in {Relation} to the {Lunar} {Reference} {System}},
	volume = {96},
	issn = {1572-9672},
	url = {https://doi.org/10.1023/A:1011989004263},
	doi = {10.1023/A:1011989004263},
	language = {en},
	number = {1},
	urldate = {2026-02-25},
	journal = {Space Science Reviews},
	author = {Neukum, G. and Ivanov, B.A. and Hartmann, W.K.},
	month = apr,
	year = {2001},
	pages = {55--86},
}

@article{hartmann_possible_2007,
	title = {Possible long-term decline in impact rates: 2. {Lunar} impact-melt data regarding impact history},
	volume = {186},
	issn = {0019-1035},
	shorttitle = {Possible long-term decline in impact rates},
	url = {https://www.sciencedirect.com/science/article/pii/S0019103506003150},
	doi = {10.1016/j.icarus.2006.09.009},
	number = {1},
	urldate = {2026-02-25},
	journal = {Icarus},
	author = {Hartmann, William K. and Quantin, Cathy and Mangold, Nicolas},
	month = jan,
	year = {2007},
	pages = {11--23},
}

@inproceedings{galluzzi_bepicolombo_2022,
	address = {Granada},
	title = {{BepiColombo} {Unique} {Views} of {Mercury} {Seven} {Years} {After} {MESSENGER}},
	url = {https://meetingorganizer.copernicus.org/EPSC2022/EPSC2022-1239.html},
	doi = {10.5194/epsc2022-1239},
	language = {en},
	number = {EPSC2022-1239},
	urldate = {2026-02-20},
	publisher = {Copernicus Meetings},
	author = {Galluzzi, Valentina and Wright, Jack and Rothery, David and Simioni, Emanuele and Zender, Joe and Benkhoff, Johannes and Cremonese, Gabriele},
	month = jul,
	year = {2022},
}

@article{rezich_investigating_2025,
	title = {Investigating the {Geotechnical} {Properties} of the {Lunar} {South} {Pole} with {NASA} {VIPER}’s {Mobility} {System}},
	volume = {6},
	issn = {2632-3338},
	url = {https://iopscience.iop.org/article/10.3847/PSJ/add13f/meta},
	doi = {10.3847/PSJ/add13f},
	language = {en},
	number = {7},
	urldate = {2026-02-20},
	journal = {The Planetary Science Journal},
	publisher = {IOP Publishing},
	author = {Rezich, Erin and Bickel, Valentin T. and Francis, Parker L. and Rogg, Arno and Tardy, Antoine and Creager, Colin and Oravec, Heather A. and Schepelmann, Alexander and Ennico-Smith, Kimberly and Deutsch, Ariel and Hirabayashi, Masatoshi},
	month = jul,
	year = {2025},
	pages = {169},
}

@article{lai_first_2020,
	title = {First look by the {Yutu}-2 rover at the deep subsurface structure at the lunar farside},
	volume = {11},
	copyright = {2020 The Author(s)},
	issn = {2041-1723},
	url = {https://www.nature.com/articles/s41467-020-17262-w},
	doi = {10.1038/s41467-020-17262-w},
	language = {en},
	number = {1},
	urldate = {2026-02-02},
	journal = {Nature Communications},
	publisher = {Nature Publishing Group},
	author = {Lai, Jialong and Xu, Yi and Bugiolacchi, Roberto and Meng, Xu and Xiao, Long and Xie, Minggang and Liu, Bin and Di, Kaichang and Zhang, Xiaoping and Zhou, Bin and Shen, Shaoxiang and Xu, Luyuan},
	month = jul,
	year = {2020},
	pages = {3426},
}

@article{warren_gprmax_2016,
	title = {{gprMax}: {Open} source software to simulate electromagnetic wave propagation for {Ground} {Penetrating} {Radar}},
	volume = {209},
	issn = {0010-4655},
	shorttitle = {{gprMax}},
	url = {https://www.sciencedirect.com/science/article/pii/S0010465516302533},
	doi = {10.1016/j.cpc.2016.08.020},
	urldate = {2026-02-03},
	journal = {Computer Physics Communications},
	author = {Warren, Craig and Giannopoulos, Antonios and Giannakis, Iraklis},
	month = dec,
	year = {2016},
	pages = {163--170},
}

@article{fah_inversion_2003,
	title = {Inversion of local {S}-wave velocity structures from average {H}/{V} ratios, and their use for the estimation of site-effects},
	volume = {7},
	issn = {1383-4649},
	doi = {10.1023/B:JOSE.0000005712.86058.42},
	number = {4},
	journal = {Journal of Seismology},
	author = {Fäh, Donat and Kind, Fortunat and Giardini, Domenico},
	year = {2003},
	pages = {449--467},
}

@article{blanchette-guertin_investigation_2012,
	title = {Investigation of scattering in lunar seismic coda},
	volume = {117},
	url = {http://doi.wiley.com/10.1029/2011JE004042},
	doi = {10.1029/2011JE004042},
	number = {E6},
	journal = {Journal of Geophysical Research},
	author = {Blanchette-Guertin, J.-F. and Johnson, Catherine L. and Lawrence, Jesse F.},
	month = jun,
	year = {2012},
	pages = {E06003--E06003},
}

@article{hobiger_shallow_2021,
	title = {The shallow structure of {Mars} at the {InSight} landing site from inversion of ambient vibrations},
	volume = {12},
	copyright = {2021 The Author(s)},
	issn = {2041-1723},
	url = {https://www.nature.com/articles/s41467-021-26957-7},
	doi = {10.1038/s41467-021-26957-7},
	language = {en},
	number = {1},
	urldate = {2021-11-23},
	journal = {Nature Communications},
	author = {Hobiger, M. and Hallo, M. and Schmelzbach, C. and Stähler, S. C. and Fäh, D. and Giardini, D. and Golombek, M. and Clinton, J. and Dahmen, N. and Zenhäusern, G. and Knapmeyer-Endrun, B. and Carrasco, S. and Charalambous, C. and Hurst, K. and Kedar, S. and Banerdt, W. B.},
	month = nov,
	year = {2021},
	pages = {6756},
}

@article{johnson_crustal_2020,
	title = {Crustal and time-varying magnetic fields at the {InSight} landing site on {Mars}},
	volume = {13},
	copyright = {2020 The Author(s), under exclusive licence to Springer Nature Limited},
	issn = {1752-0908},
	url = {https://www.nature.com/articles/s41561-020-0537-x},
	doi = {10.1038/s41561-020-0537-x},
	language = {en},
	number = {3},
	urldate = {2021-06-17},
	journal = {Nature Geoscience},
	publisher = {Nature Publishing Group},
	author = {Johnson, Catherine L. and Mittelholz, Anna and Langlais, Benoit and Russell, Christopher T. and Ansan, Véronique and Banfield, Don and Chi, Peter J. and Fillingim, Matthew O. and Forget, Francois and Haviland, Heidi Fuqua and Golombek, Matthew and Joy, Steve and Lognonné, Philippe and Liu, Xinping and Michaut, Chloé and Pan, Lu and Quantin-Nataf, Cathy and Spiga, Aymeric and Stanley, Sabine and Thorne, Shea N. and Wieczorek, Mark A. and Yu, Yanan and Smrekar, Suzanne E. and Banerdt, William B.},
	month = mar,
	year = {2020},
	note = {Number: 3},
	pages = {199--204},
}

@article{carrasco_empirical_2023,
	title = {Empirical {H}/{V} spectral ratios at the {InSight} landing site and implications for the martian subsurface structure},
	volume = {232},
	copyright = {All rights reserved},
	issn = {0956-540X},
	url = {https://doi.org/10.1093/gji/ggac391},
	doi = {10.1093/gji/ggac391},
	number = {2},
	urldate = {2022-11-29},
	journal = {Geophysical Journal International},
	author = {Carrasco, Sebastián and Knapmeyer-Endrun, Brigitte and Margerin, Ludovic and Schmelzbach, Cédric and Onodera, Keisuke and Pan, Lu and Lognonné, Philippe and Menina, Sabrina and Giardini, Domenico and Stutzmann, Eléonore and Clinton, John and Stähler, Simon and Schimmel, Martin and Golombek, Matthew and Hobiger, Manuel and Hallo, Miroslav and Kedar, Sharon and Banerdt, William Bruce},
	month = feb,
	year = {2023},
	pages = {1293--1310},
}

@article{hooper_leto_2023,
	title = {Leto {Mission} {Concept} for {Green} {Reconnaissance} of the {Marius} {Hills} {Lunar} {Pit}},
	volume = {4},
	issn = {2632-3338},
	url = {https://iopscience.iop.org/article/10.3847/PSJ/acaf87},
	doi = {10.3847/PSJ/acaf87},
	language = {en},
	number = {2},
	urldate = {2026-01-05},
	journal = {The Planetary Science Journal},
	publisher = {IOP Publishing},
	author = {Hooper, Donald M. and Ximenes, Samuel W. and Patrick, Edward L. and Wells, Ronald and Shaffer, Allison and Necsoiu, Marius},
	month = feb,
	year = {2023},
	pages = {26},
}

@article{miaja_robocrane_2022,
	title = {{RoboCrane}: {A} system for providing a power and a communication link between lunar surface and lunar caves for exploring robots},
	volume = {192},
	issn = {00945765},
	shorttitle = {{RoboCrane}},
	url = {https://linkinghub.elsevier.com/retrieve/pii/S0094576521006202},
	doi = {10.1016/j.actaastro.2021.11.023},
	language = {en},
	urldate = {2026-01-05},
	journal = {Acta Astronautica},
	author = {Miaja, Pablo F. and Navarro-Medina, Fermin and Aller, Daniel G. and León, Germán and Camanzo, Alejandro and Suarez, Carlos Manuel and Alonso, Francisco G. and Nodar, Diego and Sauro, Francesco and Bandecchi, Massimo and Bessone, Loredana and Aguado-Agelet, Fernando and Arias, Manuel},
	month = mar,
	year = {2022},
	pages = {30--46},
}

@article{losarcos_lava_2025,
	title = {Lava {Tubes} on {Earth}, the {Moon}, and {Mars}: {Detection}, {Evolution}, and {Exploration} {Potential}},
	volume = {221},
	issn = {0038-6308, 1572-9672},
	shorttitle = {Lava {Tubes} on {Earth}, the {Moon}, and {Mars}},
	url = {https://link.springer.com/10.1007/s11214-025-01260-9},
	doi = {10.1007/s11214-025-01260-9},
	language = {en},
	number = {8},
	urldate = {2025-12-17},
	journal = {Space Science Reviews},
	author = {Losarcos, Juan M. and Dombard, Andrew J.},
	month = dec,
	year = {2025},
	pages = {127},
}

@article{kaku_detection_2017,
	title = {Detection of {Intact} {Lava} {Tubes} at {Marius} {Hills} on the {Moon} by {SELENE} ({Kaguya}) {Lunar} {Radar} {Sounder}},
	volume = {44},
	copyright = {©2017. American Geophysical Union. All Rights Reserved.},
	issn = {1944-8007},
	url = {https://onlinelibrary.wiley.com/doi/abs/10.1002/2017GL074998},
	doi = {10.1002/2017GL074998},
	language = {en},
	number = {20},
	urldate = {2025-12-17},
	journal = {Geophysical Research Letters},
	author = {Kaku, T. and Haruyama, J. and Miyake, W. and Kumamoto, A. and Ishiyama, K. and Nishibori, T. and Yamamoto, K. and Crites, Sarah T. and Michikami, T. and Yokota, Y. and Sood, R. and Melosh, H. J. and Chappaz, L. and Howell, K. C.},
	year = {2017},
	note = {\_eprint: https://agupubs.onlinelibrary.wiley.com/doi/pdf/10.1002/2017GL074998},
	pages = {10,155--10,161},
}

@article{hurwitz_lunar_2013,
	title = {Lunar sinuous rilles: {Distribution}, characteristics, and implications for their origin},
	volume = {79-80},
	issn = {0032-0633},
	shorttitle = {Lunar sinuous rilles},
	url = {https://www.sciencedirect.com/science/article/pii/S0032063312003303},
	doi = {10.1016/j.pss.2012.10.019},
	urldate = {2025-12-17},
	journal = {Planetary and Space Science},
	author = {Hurwitz, Debra M. and Head, James W. and Hiesinger, Harald},
	month = may,
	year = {2013},
	pages = {1--38},
}

@article{gault_saturation_1970,
	title = {Saturation and equilibrium conditions for impact cratering on the lunar surface: {Criteria} and implications},
	volume = {5},
	issn = {1944-799X},
	shorttitle = {Saturation and equilibrium conditions for impact cratering on the lunar surface},
	url = {https://ieeexplore.ieee.org/abstract/document/7772470},
	doi = {10.1029/RS005i002p00273},
	number = {2},
	urldate = {2025-12-10},
	journal = {Radio Science},
	author = {Gault, Donald E.},
	month = feb,
	year = {1970},
	pages = {273--291},
}

@article{kolvenbach_lunarleapermission_2026,
	title = {{LunarLeaper}—{A} mission concept to explore the lunar subsurface with a small-scale legged robot},
	volume = {240},
	issn = {0094-5765},
	url = {https://www.sciencedirect.com/science/article/pii/S0094576525008082},
	doi = {10.1016/j.actaastro.2025.11.039},
	urldate = {2025-12-05},
	journal = {Acta Astronautica},
	author = {Kolvenbach, Hendrik and Mittelholz, Anna and Stähler, Simon C. and Arm, Philip and Bickel, Valentin T. and Fuhrer, Adrian and Jodar, Jordi Gomez and Margarit, Ramon and Church, Joseph and Krasnova, Elena and Walas, Krzysztof and Grott, Matthias and Hamran, Svein-Erik and Karatekin, {\textbackslash}"Ozg{\textbackslash}"ur and Olivares-Mendez, Miguel and Coloma, Sofia and Pagnamenta, Marco and Gumiela, Michał and Aaron, Jordan and Hutter, Marco},
	month = mar,
	year = {2026},
	pages = {63--75},
}

@article{fa_regolith_2015,
	title = {Regolith stratigraphy at the {Chang}'{E}-3 landing site as seen by lunar penetrating radar},
	volume = {42},
	copyright = {©2015. American Geophysical Union. All Rights Reserved.},
	issn = {1944-8007},
	url = {https://onlinelibrary.wiley.com/doi/abs/10.1002/2015GL066537},
	doi = {10.1002/2015GL066537},
	language = {en},
	number = {23},
	urldate = {2025-12-03},
	journal = {Geophysical Research Letters},
	author = {Fa, Wenzhe and Zhu, Meng-Hua and Liu, Tiantian and Plescia, Jeffrey B.},
	year = {2015},
	note = {\_eprint: https://agupubs.onlinelibrary.wiley.com/doi/pdf/10.1002/2015GL066537},
	pages = {10,179--10,187},
}

@book{heiken_lunar_1991,
	title = {Lunar sourcebook: {A} user's guide to the {Moon}},
	number = {1259},
	publisher = {Cup Archive},
	author = {Heiken, Grant and Vaniman, David and French, Bevan M},
	year = {1991},
}

@article{gong_thicknesses_2016,
	title = {Thicknesses of mare basalts on the {Moon} from gravity and topography},
	volume = {121},
	copyright = {©2016. The Authors.},
	issn = {2169-9100},
	url = {https://onlinelibrary.wiley.com/doi/abs/10.1002/2016JE005008},
	doi = {10.1002/2016JE005008},
	language = {en},
	number = {5},
	urldate = {2025-12-03},
	journal = {Journal of Geophysical Research: Planets},
	author = {Gong, Shengxia and Wieczorek, Mark A. and Nimmo, Francis and Kiefer, Walter S. and Head, James W. and Huang, Chengli and Smith, David E. and Zuber, Maria T.},
	year = {2016},
	note = {\_eprint: https://agupubs.onlinelibrary.wiley.com/doi/pdf/10.1002/2016JE005008},
	pages = {854--870},
}

@article{li_moons_2020,
	title = {The {Moon}’s farside shallow subsurface structure unveiled by {Chang}’{E}-4 {Lunar} {Penetrating} {Radar}},
	volume = {6},
	url = {https://www.science.org/doi/10.1126/sciadv.aay6898},
	doi = {10.1126/sciadv.aay6898},
	number = {9},
	urldate = {2025-12-02},
	journal = {Science Advances},
	publisher = {American Association for the Advancement of Science},
	author = {Li, Chunlai and Su, Yan and Pettinelli, Elena and Xing, Shuguo and Ding, Chunyu and Liu, Jianjun and Ren, Xin and Lauro, Sebastian E. and Soldovieri, Francesco and Zeng, Xingguo and Gao, Xingye and Chen, Wangli and Dai, Shun and Liu, Dawei and Zhang, Guangliang and Zuo, Wei and Wen, Weibin and Zhang, Zhoubin and Zhang, Xiaoxia and Zhang, Hongbo},
	month = feb,
	year = {2020},
	pages = {eaay6898},
}

@article{kiefer_gravity_2013,
	title = {Gravity constraints on the subsurface structure of the {Marius} {Hills}: {The} magmatic plumbing of the largest lunar volcanic dome complex},
	volume = {118},
	copyright = {©2012. American Geophysical Union. All Rights Reserved.},
	issn = {2169-9100},
	shorttitle = {Gravity constraints on the subsurface structure of the {Marius} {Hills}},
	url = {https://onlinelibrary.wiley.com/doi/abs/10.1029/2012JE004111},
	doi = {10.1029/2012JE004111},
	language = {en},
	number = {4},
	urldate = {2025-12-02},
	journal = {Journal of Geophysical Research: Planets},
	author = {Kiefer, Walter S.},
	year = {2013},
	note = {\_eprint: https://agupubs.onlinelibrary.wiley.com/doi/pdf/10.1029/2012JE004111},
	pages = {733--745},
}

@book{taylor_planetary_1983,
	title = {Planetary science : a lunar perspective},
	language = {en},
	publisher = {Lunar and Planetary Institute},
	author = {Taylor, Stuart Ross},
	year = {1983},
}

@article{giguere_titanium_2000,
	title = {The titanium contents of lunar mare basalts},
	volume = {35},
	issn = {1086-9379},
	url = {https://ui.adsabs.harvard.edu/abs/2000M\&PS...35..193G/abstract},
	doi = {10.1111/j.1945-5100.2000.tb01985.x},
	language = {en},
	number = {1},
	urldate = {2025-12-02},
	journal = {Meteoritics and Planetary Science},
	author = {Giguere, Thomas A. and Taylor, G. Jeffrey and Hawke, B. Ray and Lucey, Paul G.},
	month = jan,
	year = {2000},
	pages = {193--200},
}

@article{shearer_invited_1999,
	title = {{INVITED} {REVIEW}. {Magmatic} evolution of the {Moon}},
	volume = {84},
	copyright = {De Gruyter expressly reserves the right to use all content for commercial text and data mining within the meaning of Section 44b of the German Copyright Act.},
	issn = {1945-3027},
	url = {https://www.degruyterbrill.com/document/doi/10.2138/am-1999-1001/html},
	doi = {10.2138/am-1999-1001},
	language = {en},
	number = {10},
	urldate = {2025-12-01},
	journal = {American Mineralogist},
	publisher = {De Gruyter},
	author = {Shearer, C. K. and Papike, J. J.},
	month = oct,
	year = {1999},
	pages = {1469--1494},
}

@article{needham_lunar_2017,
	title = {Lunar volcanism produced a transient atmosphere around the ancient {Moon}},
	volume = {478},
	issn = {0012-821X},
	url = {https://www.sciencedirect.com/science/article/pii/S0012821X17304971},
	doi = {10.1016/j.epsl.2017.09.002},
	urldate = {2025-12-01},
	journal = {Earth and Planetary Science Letters},
	author = {Needham, Debra H. and Kring, David A.},
	month = nov,
	year = {2017},
	pages = {175--178},
}

@article{carrer_radar_2024,
	title = {Radar evidence of an accessible cave conduit on the {Moon} below the {Mare} {Tranquillitatis} pit},
	volume = {8},
	copyright = {2024 The Author(s), under exclusive licence to Springer Nature Limited},
	issn = {2397-3366},
	url = {https://www.nature.com/articles/s41550-024-02302-y},
	doi = {10.1038/s41550-024-02302-y},
	language = {en},
	number = {9},
	urldate = {2025-11-26},
	journal = {Nature Astronomy},
	publisher = {Nature Publishing Group},
	author = {Carrer, Leonardo and Pozzobon, Riccardo and Sauro, Francesco and Castelletti, Davide and Patterson, Gerald Wesley and Bruzzone, Lorenzo},
	month = sep,
	year = {2024},
	pages = {1119--1126},
}

@article{wagner_lunar_2022,
	title = {Lunar {Pit} {Morphology}: {Implications} for {Exploration}},
	volume = {127},
	copyright = {© 2022. American Geophysical Union. All Rights Reserved.},
	issn = {2169-9100},
	shorttitle = {Lunar {Pit} {Morphology}},
	url = {https://onlinelibrary.wiley.com/doi/abs/10.1029/2022JE007328},
	doi = {10.1029/2022JE007328},
	language = {en},
	number = {8},
	urldate = {2025-11-26},
	journal = {Journal of Geophysical Research: Planets},
	author = {Wagner, R. V. and Robinson, M. S.},
	year = {2022},
	note = {\_eprint: https://agupubs.onlinelibrary.wiley.com/doi/pdf/10.1029/2022JE007328},
	pages = {e2022JE007328},
}

@article{wagner_distribution_2014,
	title = {Distribution, formation mechanisms, and significance of lunar pits},
	volume = {237},
	issn = {0019-1035},
	url = {https://www.sciencedirect.com/science/article/pii/S0019103514001857},
	doi = {10.1016/j.icarus.2014.04.002},
	urldate = {2025-11-26},
	journal = {Icarus},
	author = {Wagner, Robert V. and Robinson, Mark S.},
	month = jul,
	year = {2014},
	pages = {52--60},
}

@article{ozima_terrestrial_2005,
	title = {Terrestrial nitrogen and noble gases in lunar soils},
	volume = {436},
	issn = {1476-4687},
	doi = {10.1038/nature03929},
	language = {eng},
	number = {7051},
	journal = {Nature},
	author = {Ozima, M. and Seki, K. and Terada, N. and Miura, Y. N. and Podosek, F. A. and Shinagawa, H.},
	month = aug,
	year = {2005},
	pages = {655--659},
}

@article{fagents_preservation_2010,
	title = {Preservation potential of implanted solar wind volatiles in lunar paleoregolith deposits buried by lava flows},
	volume = {207},
	issn = {0019-1035},
	url = {https://www.sciencedirect.com/science/article/pii/S0019103509004837},
	doi = {10.1016/j.icarus.2009.11.033},
	number = {2},
	urldate = {2025-12-01},
	journal = {Icarus},
	author = {Fagents, Sarah A. and Elise Rumpf, M. and Crawford, Ian A. and Joy, Katherine H.},
	month = jun,
	year = {2010},
	pages = {595--604},
}

@article{ozima_toward_2008,
	title = {Toward understanding early {Earth} evolution: prescription for approach from terrestrial noble gas and light element records in lunar soils},
	volume = {105},
	issn = {1091-6490},
	shorttitle = {Toward understanding early {Earth} evolution},
	doi = {10.1073/pnas.0806596105},
	language = {eng},
	number = {46},
	journal = {Proceedings of the National Academy of Sciences of the United States of America},
	author = {Ozima, Minoru and Yin, Qing-Zhu and Podosek, Frank A. and Miura, Yayoi N.},
	month = nov,
	year = {2008},
	pages = {17654--17658},
}

@article{murase_viscosity_1970,
	title = {Viscosity of {Lunar} {Lavas}},
	volume = {167},
	url = {https://www.science.org/doi/10.1126/science.167.3924.1491},
	doi = {10.1126/science.167.3924.1491},
	number = {3924},
	urldate = {2025-11-26},
	journal = {Science},
	publisher = {American Association for the Advancement of Science},
	author = {Murase, Tsutomu and McBirney, Alexander R.},
	month = mar,
	year = {1970},
	pages = {1491--1493},
}

@article{horvath_thermal_2022,
	title = {Thermal and {Illumination} {Environments} of {Lunar} {Pits} and {Caves}: {Models} and {Observations} {From} the {Diviner} {Lunar} {Radiometer} {Experiment}},
	volume = {49},
	copyright = {© 2022 The Authors.},
	issn = {1944-8007},
	shorttitle = {Thermal and {Illumination} {Environments} of {Lunar} {Pits} and {Caves}},
	url = {https://onlinelibrary.wiley.com/doi/abs/10.1029/2022GL099710},
	doi = {10.1029/2022GL099710},
	language = {en},
	number = {14},
	urldate = {2025-11-26},
	journal = {Geophysical Research Letters},
	author = {Horvath, Tyler and Hayne, Paul O. and Paige, David A.},
	year = {2022},
	note = {\_eprint: https://agupubs.onlinelibrary.wiley.com/doi/pdf/10.1029/2022GL099710},
	pages = {e2022GL099710},
}

@article{angelis_lunar_2002,
	title = {Lunar {Lava} {Tube} {Radiation} {Safety} {Analysis}},
	volume = {43},
	issn = {0449-3060},
	url = {https://doi.org/10.1269/jrr.43.S41},
	doi = {10.1269/jrr.43.S41},
	number = {Suppl},
	urldate = {2025-11-26},
	journal = {Journal of Radiation Research},
	author = {Angelis, De Giovanni and Wilson, J.W. and Clowdsley, M.S. and Nealy, J.E. and Humes, D.H. and Clem, J.M.},
	month = dec,
	year = {2002},
	pages = {S41--S45},
}

@article{montanez_munoz_gravity_2025,
	title = {Gravity {Modeling} of {Lunar} {Lava} {Tubes}: {Insights} {From} a {Terrestrial} {Analog}},
	volume = {130},
	copyright = {© 2025. His Majesty the King in Right of Canada and The Author(s). Reproduced with the permission of the Minister of Energy and Natural Resources.},
	issn = {2169-9100},
	shorttitle = {Gravity {Modeling} of {Lunar} {Lava} {Tubes}},
	url = {https://onlinelibrary.wiley.com/doi/abs/10.1029/2025JE009130},
	doi = {10.1029/2025JE009130},
	language = {en},
	number = {11},
	urldate = {2025-11-26},
	journal = {Journal of Geophysical Research: Planets},
	author = {Montañez Muñoz, Sarai and Williams-Jones, Glyn and Hayward, Nathan and Calahorrano-Di Patre, Antonina and Bahreyni, Behraad},
	year = {2025},
	note = {\_eprint: https://agupubs.onlinelibrary.wiley.com/doi/pdf/10.1029/2025JE009130},
	pages = {e2025JE009130},
}

@article{wang_enhanced_2025,
	title = {Enhanced {Seismic} {Backscattering} for {Lava} {Tube} {Detection}},
	volume = {52},
	copyright = {© 2025 The Author(s).},
	issn = {1944-8007},
	url = {https://onlinelibrary.wiley.com/doi/abs/10.1029/2025GL116494},
	doi = {10.1029/2025GL116494},
	language = {en},
	number = {16},
	urldate = {2025-11-24},
	journal = {Geophysical Research Letters},
	author = {Wang, Jingchuan and Schmerr, Nicholas C. and McCall, Naoma and Kruse, Sarah and Lekić, Vedran and Whelley, Patrick L. and Giles, Jacob and Wike, Linden and West, John D. and Porter, Ryan and Banks, Maria E. and Coonan, John and Deykes, Naya and Shahid, Saffat and Vig, Zachary and Zanetti, Michael},
	year = {2025},
	note = {\_eprint: https://agupubs.onlinelibrary.wiley.com/doi/pdf/10.1029/2025GL116494},
	pages = {e2025GL116494},
}

@article{torrese_detection_2021,
	title = {Detection, imaging and analysis of lava tubes for planetary analogue studies using electric methods ({ERT})},
	volume = {357},
	issn = {0019-1035},
	url = {https://www.sciencedirect.com/science/article/pii/S0019103520305728},
	doi = {10.1016/j.icarus.2020.114244},
	urldate = {2025-11-24},
	journal = {Icarus},
	author = {Torrese, Patrizio and Pozzobon, Riccardo and Rossi, Angelo Pio and Unnithan, Vikram and Sauro, Francesco and Borrmann, Dorit and Lauterbach, Helge and Santagata, Tommaso},
	month = mar,
	year = {2021},
	pages = {114244},
}

@article{torrese_planetary_2022,
	title = {Planetary analogue study using microseismic analysis for near-surface lava tube detection and exploration},
	volume = {377},
	issn = {0019-1035},
	url = {https://www.sciencedirect.com/science/article/pii/S0019103522000343},
	doi = {10.1016/j.icarus.2022.114912},
	urldate = {2025-11-24},
	journal = {Icarus},
	author = {Torrese, Patrizio and Unnithan, Vikram and Rossi, Angelo Pio},
	month = may,
	year = {2022},
	pages = {114912},
}

@article{esmaeili_resolution_2020,
	title = {Resolution of {Lava} {Tubes} {With} {Ground} {Penetrating} {Radar}: {The} {TubeX} {Project}},
	volume = {125},
	copyright = {©2020. American Geophysical Union. All Rights Reserved. This article has been contributed to by US Government employees and their work is in the public domain in the USA.},
	issn = {2169-9100},
	shorttitle = {Resolution of {Lava} {Tubes} {With} {Ground} {Penetrating} {Radar}},
	url = {https://onlinelibrary.wiley.com/doi/abs/10.1029/2019JE006138},
	doi = {10.1029/2019JE006138},
	language = {en},
	number = {5},
	urldate = {2025-11-24},
	journal = {Journal of Geophysical Research: Planets},
	author = {Esmaeili, S. and Kruse, S. and Jazayeri, S. and Whelley, P. and Bell, E. and Richardson, J. and Garry, W. B. and Young, K.},
	year = {2020},
	note = {\_eprint: https://agupubs.onlinelibrary.wiley.com/doi/pdf/10.1029/2019JE006138},
	pages = {e2019JE006138},
}

@article{mittelholz_origin_2020,
	title = {The {Origin} of {Observed} {Magnetic} {Variability} for a {Sol} on {Mars} {From} {InSight}},
	volume = {125},
	copyright = {©2020. American Geophysical Union. All Rights Reserved.},
	issn = {2169-9100},
	url = {https://onlinelibrary.wiley.com/doi/abs/10.1029/2020JE006505},
	doi = {10.1029/2020JE006505},
	language = {en},
	number = {9},
	urldate = {2025-08-28},
	journal = {Journal of Geophysical Research: Planets},
	author = {Mittelholz, A. and Johnson, C. L. and Thorne, S. N. and Joy, S. and Barrett, E. and Fillingim, M. O. and Forget, F. and Langlais, B. and Russell, C. T. and Spiga, A. and Smrekar, S. and Banerdt, W. B.},
	year = {2020},
	note = {\_eprint: https://agupubs.onlinelibrary.wiley.com/doi/pdf/10.1029/2020JE006505},
	pages = {e2020JE006505},
}

@book{carrier_d_w_olhoeft_g_r_mendell_w_physical_1991,
	title = {Physical {Properties} of the {Lunar} {Surface} - {Lunar} {Sourcebook}},
	author = {{Carrier, D. W., Olhoeft, G. R., Mendell, W.}},
	year = {1991},
}

@incollection{sauro_volcanic_2019,
	address = {Cham},
	title = {Volcanic {Caves} of {Lanzarote}: {A} {Natural} {Laboratory} for {Understanding} {Volcano}-{Speleogenetic} {Processes} and {Planetary} {Caves}},
	isbn = {978-3-030-13130-2},
	shorttitle = {Volcanic {Caves} of {Lanzarote}},
	url = {https://doi.org/10.1007/978-3-030-13130-2-9},
	doi = {10.1007/978-3-030-13130-2-9},
	language = {en},
	urldate = {2025-08-20},
	booktitle = {Lanzarote and {Chinijo} {Islands} {Geopark}: {From} {Earth} to {Space}},
	publisher = {Springer International Publishing},
	author = {Sauro, Francesco and Pozzobon, Riccardo and Santagata, Tommaso and Tomasi, Ilaria and Tonello, Matteo and Martínez-Frías, Jesús and Smets, Laurens M. Johannes and Santana Gómez, Gustavo David and Massironi, Matteo},
	editor = {Mateo, Elena and Martínez-Frías, Jesús and Vegas, Juana},
	year = {2019},
	pages = {125--142},
}

@article{arm_scientific_2023,
	title = {Scientific exploration of challenging planetary analog environments with a team of legged robots},
	volume = {8},
	url = {https://www.science.org/doi/full/10.1126/scirobotics.ade9548},
	doi = {10.1126/scirobotics.ade9548},
	number = {80},
	urldate = {2025-08-18},
	journal = {Science Robotics},
	publisher = {American Association for the Advancement of Science},
	author = {Arm, Philip and Waibel, Gabriel and Preisig, Jan and Tuna, Turcan and Zhou, Ruyi and Bickel, Valentin and Ligeza, Gabriela and Miki, Takahiro and Kehl, Florian and Kolvenbach, Hendrik and Hutter, Marco},
	month = jul,
	year = {2023},
	pages = {eade9548},
}

@article{sauro_lava_2020,
	title = {Lava tubes on {Earth}, {Moon} and {Mars}: {A} review on their size and morphology revealed by comparative planetology},
	volume = {209},
	issn = {0012-8252},
	shorttitle = {Lava tubes on {Earth}, {Moon} and {Mars}},
	url = {https://www.sciencedirect.com/science/article/pii/S0012825220303342},
	doi = {10.1016/j.earscirev.2020.103288},
	urldate = {2025-08-18},
	journal = {Earth-Science Reviews},
	author = {Sauro, Francesco and Pozzobon, Riccardo and Massironi, Matteo and De Berardinis, Pierluigi and Santagata, Tommaso and De Waele, Jo},
	month = oct,
	year = {2020},
	pages = {103288},
}

@article{keszthelyi_physical_1998,
	title = {Some physical requirements for the emplacement of long basaltic lava flows},
	volume = {103},
	issn = {2156-2202},
	url = {https://onlinelibrary.wiley.com/doi/abs/10.1029/98JB00606},
	doi = {10.1029/98JB00606},
	language = {en},
	number = {B11},
	urldate = {2025-08-18},
	journal = {Journal of Geophysical Research: Solid Earth},
	author = {Keszthelyi, L. and Self, S.},
	year = {1998},
	note = {\_eprint: https://agupubs.onlinelibrary.wiley.com/doi/pdf/10.1029/98JB00606},
	pages = {27447--27464},
}

@article{zhu_grail_2024,
	title = {{GRAIL} gravity gradients evidence for a potential lava tube at {Marius} {Hills} on the moon},
	volume = {408},
	issn = {0019-1035},
	url = {https://www.sciencedirect.com/science/article/pii/S0019103523003937},
	doi = {10.1016/j.icarus.2023.115814},
	urldate = {2025-08-18},
	journal = {Icarus},
	author = {Zhu, Ke and Yang, Meng and Yan, XingYuan and Li, WeiKai and Feng, Wei and Zhong, Min},
	month = jan,
	year = {2024},
	pages = {115814},
}

@techreport{morgavi_tubes_2025,
	title = {{TUBES}: a multidisciplinary project for understanding lava tube formation and preservation at {Vesuvius} and {Etna}.\&nbsp;},
	shorttitle = {{TUBES}},
	url = {https://meetingorganizer.copernicus.org/EGU25/EGU25-13496.html},
	doi = {10.5194/egusphere-egu25-13496},
	language = {en},
	number = {EGU25-13496},
	urldate = {2025-08-18},
	institution = {Copernicus Meetings},
	author = {Morgavi, Daniele and Calvari, Sonia and Barile, Claudia and Lemaire, Thomas and Petrosino, Paola and Martire, Diego Di and Valente, Ettore and Repola, Leopoldo and Spampinato, Letizia and Miraglia, Lucia and Macedonio, Giovanni and Giudicepietro, Flora and Pappalettera, Giovanni and Mpoyi, Dany katamba},
	month = mar,
	year = {2025},
	note = {Conference Name: EGU25},
}

@article{haruyama_possible_2009,
	title = {Possible lunar lava tube skylight observed by {SELENE} cameras},
	volume = {36},
	copyright = {Copyright 2009 by the American Geophysical Union.},
	issn = {1944-8007},
	url = {https://onlinelibrary.wiley.com/doi/abs/10.1029/2009GL040635},
	doi = {10.1029/2009GL040635},
	language = {en},
	number = {21},
	urldate = {2025-08-18},
	journal = {Geophysical Research Letters},
	author = {Haruyama, Junichi and Hioki, Kazuyuki and Shirao, Motomaro and Morota, Tomokatsu and Hiesinger, Harald and van der Bogert, Carolyn H. and Miyamoto, Hideaki and Iwasaki, Akira and Yokota, Yasuhiro and Ohtake, Makiko and Matsunaga, Tsuneo and Hara, Seiichi and Nakanotani, Shunsuke and Pieters, Carle M.},
	year = {2009},
	note = {\_eprint: https://agupubs.onlinelibrary.wiley.com/doi/pdf/10.1029/2009GL040635},
}

@article{dietterich_benchmarking_2017,
	title = {Benchmarking computational fluid dynamics models of lava flow simulation for hazard assessment, forecasting, and risk management},
	volume = {6},
	issn = {2191-5040},
	url = {https://doi.org/10.1186/s13617-017-0061-x},
	doi = {10.1186/s13617-017-0061-x},
	language = {en},
	number = {1},
	urldate = {2025-08-18},
	journal = {Journal of Applied Volcanology},
	author = {Dietterich, Hannah R. and Lev, Einat and Chen, Jiangzhi and Richardson, Jacob A. and Cashman, Katharine V.},
	month = may,
	year = {2017},
	pages = {9},
}

@article{chappaz_evidence_2017,
	title = {Evidence of large empty lava tubes on the {Moon} using {GRAIL} gravity},
	volume = {44},
	copyright = {©2016. American Geophysical Union. All Rights Reserved.},
	issn = {1944-8007},
	url = {https://onlinelibrary.wiley.com/doi/abs/10.1002/2016GL071588},
	doi = {10.1002/2016GL071588},
	language = {en},
	number = {1},
	urldate = {2025-08-18},
	journal = {Geophysical Research Letters},
	author = {Chappaz, Loic and Sood, Rohan and Melosh, Henry J. and Howell, Kathleen C. and Blair, David M. and Milbury, Colleen and Zuber, Maria T.},
	year = {2017},
	note = {\_eprint: https://agupubs.onlinelibrary.wiley.com/doi/pdf/10.1002/2016GL071588},
	pages = {105--112},
}

@article{porcello_apollo_1974,
	title = {The {Apollo} lunar sounder radar system},
	volume = {62},
	issn = {1558-2256},
	url = {https://ieeexplore.ieee.org/abstract/document/1451447},
	doi = {10.1109/PROC.1974.9517},
	number = {6},
	urldate = {2025-08-18},
	journal = {Proceedings of the IEEE},
	author = {Porcello, L.J. and Jordan, R.L. and Zelenka, J.S. and Adams, G.F. and Phillips, R.J. and Brown, W.E. and Ward, S.H. and Jackson, P.L.},
	month = jun,
	year = {1974},
	pages = {769--783},
}

@article{xu_lunar_2024,
	title = {Lunar {Exploration} {Based} on {Ground}-{Based} {Radar}: {Current} {Research} {Progress} and {Future} {Prospects}},
	volume = {16},
	copyright = {https://creativecommons.org/licenses/by/4.0/},
	issn = {2072-4292},
	shorttitle = {Lunar {Exploration} {Based} on {Ground}-{Based} {Radar}},
	url = {https://www.mdpi.com/2072-4292/16/18/3484},
	doi = {10.3390/rs16183484},
	language = {en},
	number = {18},
	urldate = {2025-08-18},
	journal = {Remote Sensing},
	author = {Xu, Jiangwan and Ding, Chunyu and Su, Yan and Ding, Zonghua and Yang, Song and Li, Jiawei and Dong, Zehua and Sharma, Ravi and Qiu, Xiaohang and Lei, Zhonghan and Chen, Haoyu and Jiang, Changzhi and Chen, Wentao and Cheng, Qi and Liang, Zihang},
	month = sep,
	year = {2024},
	pages = {3484},
}

@article{ding_moon-based_2023,
	title = {Moon-{Based} {Ground}-{Penetrating} {Radar} {Observation} of the {Latest} {Volcanic} {Activity} at the {Chang}’{E}-4 {Landing} {Site}},
	volume = {61},
	copyright = {https://ieeexplore.ieee.org/Xplorehelp/downloads/license-information/IEEE.html},
	issn = {0196-2892, 1558-0644},
	url = {https://ieeexplore.ieee.org/document/10129948/},
	doi = {10.1109/TGRS.2023.3277992},
	language = {en},
	urldate = {2025-08-18},
	journal = {IEEE Transactions on Geoscience and Remote Sensing},
	author = {Ding, Chunyu and Li, Jing and Hu, Rong},
	year = {2023},
	pages = {1--10},
}

@article{zhao_failure_2022,
	title = {Failure modes and excavation stability of large-scale columnar jointed rock masses containing interlayer shear weakness zones},
	volume = {159},
	issn = {1365-1609},
	url = {https://www.sciencedirect.com/science/article/pii/S1365160922001885},
	doi = {10.1016/j.ijrmms.2022.105222},
	urldate = {2025-08-18},
	journal = {International Journal of Rock Mechanics and Mining Sciences},
	author = {Zhao, Danchen and Xia, Yingjie and Zhang, Chuanqing and Tang, Chun'an and Zhou, Hui and Liu, Ning and Singh, Hemant Kumar and Zhao, Zhenxing and Chen, Jun and Mu, Chaoqian},
	month = nov,
	year = {2022},
	pages = {105222},
}

@article{pettinelli_overview_2022,
	title = {An overview of {GPR} subsurface exploration of planets and moons},
	volume = {41},
	issn = {1070-485X, 1938-3789},
	url = {https://library.seg.org/doi/10.1190/tle41100672.1},
	doi = {10.1190/tle41100672.1},
	language = {en},
	number = {10},
	urldate = {2025-08-18},
	journal = {The Leading Edge},
	author = {Pettinelli, Elena and Cosciotti, Barbara and Lauro, Sebastian Emanuel and Mattei, Elisabetta},
	month = oct,
	year = {2022},
	pages = {672--680},
}

@article{mittelholz_lunarleaper_2024,
	title = {Lunarleaper - {Unlocking} a {Subsurface} {World}},
	url = {https://ui.adsabs.harvard.edu/abs/2024EGUGA..2621578M/abstract},
	doi = {10.5194/egusphere-egu24-21578},
	language = {en},
	urldate = {2025-08-18},
	journal = {European Geosciences Union General Assembly 2024 (EGU24)},
	author = {Mittelholz, Anna and Stähler, Simon C. and Kolvenbach, Hendrik and Bickel, Valentin and Church, Joseph and Hamran, Svein-Erik and Karatekin, Ozgur and Ritter, Birgit and Aaron, Jordan and Anhorn, Barthélémy and Coloma, Sofia and de Palézieux dit Falconnet, Larissa and Grott, Matthias and Ogier, Cristophe and Robertsson, Johan and Walas, Krzysrof},
	month = apr,
	year = {2024},
	pages = {21578},
}

@incollection{mckay_properties_1990,
	series = {Soil {Micro}-{Morphology}: {A} {Basic} and {Applied} {Science}},
	title = {Properties of {Lunar} {Regolith}},
	volume = {19},
	url = {https://www.sciencedirect.com/science/article/pii/S016624810870360X},
	doi = {10.1016/S0166-2481(08)70360-X},
	urldate = {2025-08-18},
	booktitle = {Developments in {Soil} {Science}},
	publisher = {Elsevier},
	author = {Mckay, D. S. and Ming, D. W.},
	editor = {Douglas, Lowell A.},
	month = jan,
	year = {1990},
	pages = {449--462},
}

@article{theinat_lunar_2020,
	title = {Lunar lava tubes: {Morphology} to structural stability},
	volume = {338},
	issn = {0019-1035},
	shorttitle = {Lunar lava tubes},
	url = {https://www.sciencedirect.com/science/article/pii/S0019103518307826},
	doi = {10.1016/j.icarus.2019.113442},
	urldate = {2025-08-18},
	journal = {Icarus},
	author = {Theinat, Audai K. and Modiriasari, Anahita and Bobet, Antonio and Melosh, H. Jay and Dyke, Shirley J. and Ramirez, Julio and Maghareh, Amin and Gomez, Daniel},
	month = mar,
	year = {2020},
	pages = {113442},
}

@article{chwala_structural_2024,
	title = {Structural stability of lunar lava tubes with consideration of variable cross-section geometry},
	volume = {411},
	issn = {0019-1035},
	url = {https://www.sciencedirect.com/science/article/pii/S0019103523005079},
	doi = {10.1016/j.icarus.2023.115928},
	urldate = {2025-08-18},
	journal = {Icarus},
	author = {Chwała, Marcin and Komatsu, Goro and Haruyama, Junichi},
	month = mar,
	year = {2024},
	pages = {115928},
}

@article{bickel_analysis_2019,
	title = {Analysis of {Lunar} {Boulder} {Tracks}: {Implications} for {Trafficability} of {Pyroclastic} {Deposits}},
	volume = {124},
	copyright = {©2019. American Geophysical Union. All Rights Reserved.},
	issn = {2169-9100},
	shorttitle = {Analysis of {Lunar} {Boulder} {Tracks}},
	url = {https://onlinelibrary.wiley.com/doi/abs/10.1029/2018JE005876},
	doi = {10.1029/2018JE005876},
	language = {en},
	number = {5},
	urldate = {2025-08-18},
	journal = {Journal of Geophysical Research: Planets},
	author = {Bickel, V. T. and Honniball, C. I. and Martinez, S. N. and Rogaski, A. and Sargeant, H. M. and Bell, S. K. and Czaplinski, E. C. and Farrant, B. E. and Harrington, E. M. and Tolometti, G. D. and Kring, D. A.},
	year = {2019},
	note = {\_eprint: https://agupubs.onlinelibrary.wiley.com/doi/pdf/10.1029/2018JE005876},
	pages = {1296--1314},
}

@article{kokelaar_granular_2017,
	title = {Granular avalanches on the {Moon}: {Mass}-wasting conditions, processes, and features},
	volume = {122},
	copyright = {©2017. The Authors.},
	issn = {2169-9100},
	shorttitle = {Granular avalanches on the {Moon}},
	url = {https://onlinelibrary.wiley.com/doi/abs/10.1002/2017JE005320},
	doi = {10.1002/2017JE005320},
	language = {en},
	number = {9},
	urldate = {2025-08-18},
	journal = {Journal of Geophysical Research: Planets},
	author = {Kokelaar, B. P. and Bahia, R. S. and Joy, K. H. and Viroulet, S. and Gray, J. M. N. T.},
	year = {2017},
	note = {\_eprint: https://agupubs.onlinelibrary.wiley.com/doi/pdf/10.1002/2017JE005320},
	pages = {1893--1925},
}

@article{bickel_global_2022,
	title = {A {Global} {Perspective} on {Lunar} {Granular} {Flows}},
	volume = {49},
	copyright = {© 2022 The Authors.},
	issn = {1944-8007},
	url = {https://onlinelibrary.wiley.com/doi/abs/10.1029/2022GL098812},
	doi = {10.1029/2022GL098812},
	language = {en},
	number = {12},
	urldate = {2025-08-18},
	journal = {Geophysical Research Letters},
	author = {Bickel, V. T. and Loew, S. and Aaron, J. and Goedhart, N.},
	year = {2022},
	note = {\_eprint: https://agupubs.onlinelibrary.wiley.com/doi/pdf/10.1029/2022GL098812},
	pages = {e2022GL098812},
}

@article{hungr_varnes_2014,
	title = {The {Varnes} classification of landslide types, an update},
	volume = {11},
	issn = {1612-5118},
	url = {https://doi.org/10.1007/s10346-013-0436-y},
	doi = {10.1007/s10346-013-0436-y},
	language = {en},
	number = {2},
	urldate = {2025-08-18},
	journal = {Landslides},
	author = {Hungr, Oldrich and Leroueil, Serge and Picarelli, Luciano},
	month = apr,
	year = {2014},
	pages = {167--194},
}

@article{Williams2025,
author = {Williams, E. A. and Montési, L. G. J.},
title = {Crack Development Inside and Around Lunar Lava Tubes},
journal = {Journal of Geophysical Research: Planets},
volume = {130},
number = {3},
pages = {e2024JE008553},
doi = {https://doi.org/10.1029/2024JE008553},
url = {https://agupubs.onlinelibrary.wiley.com/doi/abs/10.1029/2024JE008553},
eprint = {https://agupubs.onlinelibrary.wiley.com/doi/pdf/10.1029/2024JE008553},
note = {e2024JE008553 2024JE008553},
year = {2025}
}

@article{keil_evaluating_2026,
  title = {Evaluating {{Seismic Ambient Noise Techniques}} for {{Imaging Lava Tubes}} on the {{Moon}}},
  author = {Keil, Sabrina and Sesko, Rok and Schimmel, Martin and Igel, Heiner and Reiss, Philipp},
  year = 2026,
  journal = {Journal of Geophysical Research: Planets},
  volume = {131},
  number = {4},
  pages = {e2025JE009576},
  issn = {2169-9100},
  doi = {10.1029/2025JE009576},
  urldate = {2026-04-09},
  copyright = {\copyright{} 2026 The Author(s).},
  langid = {english}
}

@article{lewis_surface_2019,
  title = {A Surface Gravity Traverse on {{Mars}} Indicates Low Bedrock Density at {{Gale}} Crater},
  author = {Lewis, Kevin W. and Peters, Stephen and Gonter, Kurt and Morrison, Shaunna and Schmerr, Nicholas C. and Vasavada, Ashwin R. and Gabriel, Travis},
  year = 2019,
  journal = {Science},
  volume = {363},
  number = {6426},
  pages = {535--537},
  doi = {10.1126/science.aat0738}
}

@article{noeker_design_2025,
  title = {Design, Manufacturing, and Assembly of the {{GRASS}} Small Body Gravimeter Spring},
  author = {Noeker, Matthias and Ritter, Birgit and Van Ransbeeck, Emiel and Karatekin, {\"O}zg{\"u}r},
  year = 2025,
  month = sep,
  journal = {CEAS Space Journal},
  volume = {17},
  number = {5},
  pages = {755--768},
  issn = {1868-2510},
  doi = {10.1007/s12567-024-00574-8},
  urldate = {2025-12-01},
  langid = {english}
}

@article{civilini_thermal_2023,
  title = {Thermal {{Moonquake Characterization}} and {{Cataloging Using Frequency-Based Algorithms}} and {{Stochastic Gradient Descent}}},
  author = {Civilini, F. and Weber, R. and Husker, A.},
  year = 2023,
  journal = {Journal of Geophysical Research: Planets},
  volume = {128},
  number = {9},
  pages = {e2022JE007704},
  issn = {2169-9100},
  doi = {10.1029/2022JE007704},
  urldate = {2023-09-13},
  copyright = {\copyright{} 2023 American Geophysical Union. All Rights Reserved. This article has been contributed to by U.S. Government employees and their work is in the public domain in the USA.},
  langid = {english}
}

@article{lucey_lunar_2000,
  title = {Lunar Iron and Titanium Abundance Algorithms Based on Final Processing of {{Clementine}} Ultraviolet-Visible Images},
  author = {Lucey, Paul G. and Blewett, David T. and Jolliff, Bradley L.},
  year = 2000,
  journal = {Journal of Geophysical Research: Planets},
  volume = {105},
  number = {E8},
  pages = {20297--20305},
  issn = {2156-2202},
  doi = {10.1029/1999JE001117},
  urldate = {2026-03-02},
  copyright = {Copyright 2000 by the American Geophysical Union.},
  langid = {english}
}

@article{speyerer_precise_2023,
  title = {Precise Mapping of the {{Moon}} with the {{Clementine Ultraviolet}}/{{Visible Camera}}},
  author = {Speyerer, Emerson J. and Robinson, Mark S. and Boyd, Aaron and Silva, Victor H. and Lawrence, Samuel},
  year = 2023,
  month = jul,
  journal = {Icarus},
  volume = {398},
  pages = {115506},
  issn = {0019-1035},
  doi = {10.1016/j.icarus.2023.115506},
  urldate = {2026-03-02}
}

@INPROCEEDINGS{Arm2024,
  author={Arm, Philip and Mittal, Mayank and Kolvenbach, Hendrik and Hutter, Marco},
  booktitle={2024 IEEE International Conference on Robotics and Automation (ICRA)}, 
  title={Pedipulate: Enabling Manipulation Skills using a Quadruped Robot’s Leg}, 
  year={2024},
  volume={},
  number={},
  pages={5717-5723},
  doi={10.1109/ICRA57147.2024.10611307}}

@article{Rezich2025,
doi = {10.3847/PSJ/add13f},
url = {https://doi.org/10.3847/PSJ/add13f},
year = {2025},
month = {jul},
publisher = {The American Astronomical Society},
volume = {6},
number = {7},
pages = {169},
author = {Rezich, Erin and Bickel, Valentin T. and Francis, Parker L. and Rogg, Arno and Tardy, Antoine and Creager, Colin and Oravec, Heather A. and Schepelmann, Alexander and Ennico-Smith, Kimberly and Deutsch, Ariel and Hirabayashi, Masatoshi},
title = {Investigating the Geotechnical Properties of the Lunar South Pole with NASA VIPER’s Mobility System},
journal = {The Planetary Science Journal}
}

@ARTICLE{Kolvenbach2019,
  author={Kolvenbach, Hendrik and Bärtschi, Christian and Wellhausen, Lorenz and Grandia, Ruben and Hutter, Marco},
  journal={IEEE Robotics and Automation Letters}, 
  title={Haptic Inspection of Planetary Soils With Legged Robots}, 
  year={2019},
  volume={4},
  number={2},
  pages={1626-1632},
  doi={10.1109/LRA.2019.2896732}}

@article{Sargeant_bouldertracks_2020,
author = {Sargeant, H. M. and Bickel, V. T. and Honniball, C. I. and Martinez, S. N. and Rogaski, A. and Bell, S. K. and Czaplinski, E. C. and Farrant, B. E. and Harrington, E. M. and Tolometti, G. D. and Kring, D. A.},
title = {Using Boulder Tracks as a Tool to Understand the Bearing Capacity of Permanently Shadowed Regions of the Moon},
journal = {Journal of Geophysical Research: Planets},
volume = {125},
number = {2},
pages = {e2019JE006157},
doi = {https://doi.org/10.1029/2019JE006157},
url = {https://agupubs.onlinelibrary.wiley.com/doi/abs/10.1029/2019JE006157},
eprint = {https://agupubs.onlinelibrary.wiley.com/doi/pdf/10.1029/2019JE006157},
note = {e2019JE006157 2019JE006157},
year = {2020}
}

@article{Bickel_Kring2020,
title = {Lunar south pole boulders and boulder tracks: Implications for crew and rover traverses},
journal = {Icarus},
volume = {348},
pages = {113850},
year = {2020},
issn = {0019-1035},
doi = {https://doi.org/10.1016/j.icarus.2020.113850},
url = {https://www.sciencedirect.com/science/article/pii/S0019103520302311},
author = {V.T. Bickel and D.A. Kring}
}

@inproceedings{carroll2015exploring,
  title={Exploring lunar sub-surface voids using surface gravimetry},
  author={Carroll, Kieran A and Hatch, David and Ghent, Rebecca and Stanley, Sabine and Urbancic, Natasha and Williamson, Marie-Claude and Garry, W Brent and Talwani, Manik},
  booktitle={46th Lunar and Planetary Science Conference},
  year={2015}
}

@INPROCEEDINGS{Margarit2025,
       author = {{Margarit}, Ramon and {Mittelholz}, Anna and {Bickel}, Valentin and {St{\"a}hler}, Simon},
        title = "{Geomorphometric Assessment of the Marius Hill's Pit for LunarLeaper}",
    booktitle = {EGU General Assembly Conference Abstracts},
         year = 2025,
       series = {EGU General Assembly Conference Abstracts},
        month = apr,
          eid = {EGU25-75},
        pages = {EGU25-75},
          doi = {10.5194/egusphere-egu25-75},
       adsurl = {https://ui.adsabs.harvard.edu/abs/2025EGUGA..27...75M}
}

@inproceedings{lee2018possible,
  title={Possible lava tube skylights near the North Pole of the Moon},
  author={Lee, P},
  booktitle={49th Annual Lunar and Planetary Science Conference},
  number={2083},
  pages={2982},
  year={2018}
}

@techreport{Wilhelms1987,
  title={The geologic history of the Moon},
  author={Wilhelms, Don E and McCauley, John F and Trask, Newell J},
  year={1987}
}

@article{Tarduno2021,
author = {Tarduno, John A and Cottrell, Rory D and Lawrence, Kristin and Bono, Richard K and Huang, Wentao and Johnson, Catherine L and Blackman, Eric G and Smirnov, Aleksey V and Nakajima, Miki and Neal, Clive R and Zhou, Tinghong and Ibanez-Mejia, Mauricio and Oda, Hirokuni and Crummins, Ben},
doi = {10.1126/sciadv.abi7647},
journal = {Science Advances},
number = {32},
pages = {eabi7647},
title = {{Absence of a long-lived lunar paleomagnetosphere}},
volume = {7},
year = {2021}
}

@article{Tikoo2017,
author = {Tikoo, Sonia M and Weiss, Benjamin P and Shuster, David L and Suavet, Cl{\'{e}}ment and Wang, Huapei and Grove, Timothy L},
doi = {10.1126/sciadv.1700207},
issn = {23752548},
journal = {Science Advances},
number = {8},
pages = {1--10},
pmid = {28808679},
title = {{A two-billion-year history for the lunar dynamo}},
volume = {3},
year = {2017}
}

@article{Mittelholz2025,
author = {Mittelholz, Anna and Moorkamp, Max and Broquet, Adrien and Ojha, Lujendra},
doi = {https://doi.org/10.1029/2024JE008832},
journal = {Journal of Geophysical Research: Planets},
number = {4},
pages = {e2024JE008832},
title = {{Gravity and Magnetic Field Signatures in Hydrothermally Affected Regions on Mars}},
url = {https://agupubs.onlinelibrary.wiley.com/doi/abs/10.1029/2024JE008832},
volume = {130},
year = {2025}
}

@article{cruikshank_lunar_1972,
    title = {Lunar rilles and {Hawaiian} volcanic features: {Possible} analogues},
    volume = {3},
    issn = {1573-0794},
    shorttitle = {Lunar rilles and {Hawaiian} volcanic features},
    url = {https://doi.org/10.1007/BF00562463},
    doi = {10.1007/BF00562463},
    language = {en},
    number = {4},
    urldate = {2023-12-11},
    journal = {The moon},
    author = {Cruikshank, D. P. and Wood, C. A.},
    month = mar,
    year = {1972},
    pages = {412--447},
}

@article{greeley_lava_1971,
    title = {Lava tubes and channels in the lunar {Marius} {Hills}},
    volume = {3},
    issn = {1573-0794},
    url = {https://doi.org/10.1007/BF00561842},
    doi = {10.1007/BF00561842},
    language = {en},
    number = {3},
    urldate = {2023-12-11},
    journal = {The moon},
    author = {Greeley, Ronald},
    month = dec,
    year = {1971},
    pages = {289--314},
}

@article{robinson_confirmation_2012,
    title = {Confirmation of sublunarean voids and thin layering in mare deposits},
    volume = {69},
    issn = {0032-0633},
    url = {https://www.sciencedirect.com/science/article/pii/S0032063312001195},
    doi = {10.1016/j.pss.2012.05.008},
    number = {1},
    urldate = {2023-11-16},
    journal = {Planetary and Space Science},
    author = {Robinson, M. S. and Ashley, J. W. and Boyd, A. K. and Wagner, R. V. and Speyerer, E. J. and Ray Hawke, B. and Hiesinger, H. and van der Bogert, C. H.},
    month = aug,
    year = {2012},
    pages = {18--27},
}

@inproceedings{Wolter:2025,
    author = {David Wolter and Matthias Grott and J{\"o}rg Knollenberg and Lynn Miller and Christian Althaus and Toni Gro{\ss}mann and Julia Wecker and J{\"o}rg Martin and Andreas Ihring and Boris Jung and Konstantinos Vasiliou and Benjamin Fechtig},
    title = {{Compact MEMS-based Fabry-Perot interferometers for space applications}},
    volume = {13612},
    booktitle = {Infrared Remote Sensing and Instrumentation XXXIII},
    organization = {International Society for Optics and Photonics},
    publisher = {SPIE},
    pages = {136120B},
    year = {2025},
    doi = {10.1117/12.3063608},
}

@ARTICLE{Grott:2017,
       author = {{Grott}, M. and {Knollenberg}, J. and {Borgs}, B. and {H{\"a}nschke}, F. and {Kessler}, E. and {Helbert}, J. and {Maturilli}, A. and {M{\"u}ller}, N.},
        title = "{The MASCOT Radiometer MARA for the Hayabusa 2 Mission}",
      journal = {Space Science Reviews},
         year = 2017,
        month = jul,
       volume = {208},
       number = {1-4},
        pages = {413-431},
          doi = {10.1007/s11214-016-0272-1},
}

@ARTICLE{Spohn:2018,
       author = {{Spohn}, T. and {Grott}, M. and {Smrekar}, S.~E. and {Knollenberg}, J. and {Hudson}, T.~L. and {Krause}, C. and {M{\"u}ller}, N. and {J{\"a}nchen}, J. and {B{\"o}rner}, A. and {Wippermann}, T. and {Kr{\"o}mer}, O. and {Lichtenheldt}, R. and {Wisniewski}, L. and {Grygorczuk}, J. and {Fittock}, M. and {Rheershemius}, S. and {Spr{\"o}witz}, T. and {Kopp}, E. and {Walter}, I. and {Plesa}, A.~C. and {Breuer}, D. and {Morgan}, P. and {Banerdt}, W.~B.},
        title = "{The Heat Flow and Physical Properties Package (HP$^{3}$) for the InSight Mission}",
      journal = {Space Science Reviews},
         year = 2018,
        month = aug,
       volume = {214},
       number = {5},
          eid = {96},
        pages = {96},
          doi = {10.1007/s11214-018-0531-4},
}

@inproceedings{Kessler:2005,
       author = {E. Kessler},
    booktitle = {Proc. of Sensor 2005, 12th International Conference, Nürnberg},
         year = 2005,
       volume = {Vol I},
       number = {},
        pages = {73-78}
}

@article{Shirley:2019,
    author = {Shirley, K. A. and Glotch, T. D.},
    title = {Particle Size Effects on Mid-Infrared Spectra of Lunar Analog Minerals in a Simulated Lunar Environment},
    journal = {Journal of Geophysical Research: Planets},
    volume = {124},
    number = {4},
    pages = {970-988},
    year = {2019}
}

@article{Lucy:2021,
    author = {Lucey, Paul G. and Greenhagen, Benjamin and Donaldson Hanna, Kerri and Bowles, Neil and Flom, Abigail and Paige, David A.},
    title = {Christiansen Feature Map From the Lunar Reconnaissance Orbiter Diviner Lunar Radiometer Experiment: Improved Corrections and Derived Mineralogy},
    journal = {Journal of Geophysical Research: Planets},
    volume = {126},
    number = {6},
    pages = {e2020JE006777},
    doi = {https://doi.org/10.1029/2020JE006777},
    year = {2021}
}

@BOOK{Heiken:1991,
       author = {{Heiken}, Grant H. and {Vaniman}, David T. and {French}, Bevan M.},
        title = "{Lunar Sourcebook, A User's Guide to the Moon}",
         year = 1991,
}

@article{Besse:2011,
    author = {Besse, S. and Sunshine, J. M. and Staid, M. I. and Petro, N. E. and Boardman, J. W. and Green, R. O. and Head, J. W. and Isaacson, P. J. and Mustard, J. F. and Pieters, C. M.},
    title = {Compositional variability of the Marius Hills volcanic complex from the Moon Mineralogy Mapper (M3)},
    journal = {Journal of Geophysical Research: Planets},
    volume = {116},
    number = {E6},
    pages = {},
    doi = {https://doi.org/10.1029/2010JE003725},
    year = {2011}
}

@ARTICLE{Lee:2022,
       author = {{Lee}, Sujeong and {van Riessen}, Arie},
        title = "{A Review on Geopolymer Technology for Lunar Base Construction}",
      journal = {Materials},
         year = 2022,
        month = jun,
       volume = {15},
       number = {13},
        pages = {4516},
          doi = {10.3390/ma15134516}
}

@article{Zheng:2024,
    author = {Zheng, X. and Zhao, C. and Sun, X. and Dong, W.},
    title = {Lunar Regolith Geopolymer Concrete for In-Situ Construction of Lunar Bases: A Review},
    journal = {Polymers},
    volume = {16},
    number = {},
    pages = {1582},
    year = {2024},
    doi = {10.3390/polym1611158210.34133/space.0037}
}

@article{Sargeant:2020,
title = {Hydrogen reduction of ilmenite: Towards an in situ resource utilization demonstration on the surface of the Moon},
journal = {Planetary and Space Science},
volume = {180},
pages = {104751},
year = {2020},
issn = {0032-0633},
doi = {https://doi.org/10.1016/j.pss.2019.104751},
author = {H.M. Sargeant and F.A.J. Abernethy and S.J. Barber and I.P. Wright and M. Anand and S. Sheridan and A. Morse}
}

@ARTICLE{Sargeant:2021,
       author = {{Sargeant}, H.~M. and {Barber}, S.~J. and {Anand}, M. and {Abernethy}, F.~A.~J. and {Sheridan}, S. and {Wright}, I.~P. and {Morse}, A.~D.},
        title = "{Hydrogen reduction of lunar samples in a static system for a water production demonstration on the Moon}",
      journal = {Plan. Space Sci.},
         year = 2021,
        month = oct,
       volume = {205},
          eid = {105287},
        pages = {105287},
          doi = {10.1016/j.pss.2021.105287}
}

@article{Neumann:2008,
    author = {Norbert Neumann and Martin Ebermann and Steffen Kurth and Karla Hiller},
    title = {{Tunable infrared detector with integrated micromachined Fabry-Perot filter}},
    volume = {7},
    journal = {Journal of Micro/Nanolithography, MEMS, and MOEMS},
    number = {2},
    publisher = {SPIE},
    pages = {021004},
    year = {2008},
    doi = {10.1117/1.2909206},
}

@inproceedings{Ebermann:2016,
    author = {Martin Ebermann and Norbert Neumann and Karla Hiller and Mario Seifert and Marco Meinig and Steffen Kurth},
    title = {{Tunable MEMS Fabry-Pérot filters for infrared microspectrometers: a review}},
    volume = {9760},
    booktitle = {MOEMS and Miniaturized Systems XV},
    editor = {Wibool Piyawattanametha and Yong-Hwa Park},
    organization = {International Society for Optics and Photonics},
    publisher = {SPIE},
    pages = {97600H},
    year = {2016},
    doi = {10.1117/12.2209288},
}

@inproceedings{Meining:2011,
    author = {Marco Meinig and Steffen Kurth and Karla Hiller and Norbert Neumann and Martin Ebermann and Elvira Gittler and Thomas Gessner},
    title = {{Tunable mid-infrared filter based on Fabry-Perot interferometer with two movable reflectors}},
    volume = {7930},
    booktitle = {MOEMS and Miniaturized Systems X},
    editor = {Harald Schenk and Wibool Piyawattanametha},
    organization = {International Society for Optics and Photonics},
    publisher = {SPIE},
    pages = {79300K},
    year = {2011},
    doi = {10.1117/12.875428},
}

@ARTICLE{Knollenberg:2025,
       author = {{Knollenberg}, J. and {Grott}, M. and {Hamm}, M. and {Ihring}, A. and {Ziese}, R. and {Biele}, J.},
        title = "{The miniRAD instrument for the MMX IDEFIX rover}",
      journal = {Progress in Earth and Planetary Science},
         year = 2025,
        month = jul,
       volume = {12},
       number = {1},
          eid = {53},
        pages = {53},
          doi = {10.1186/s40645-025-00717-3},
}

@article{Deroussi:2009,
title = {Localization of cavities in a thick lava flow by microgravimetry},
journal = {Journal of Volcanology and Geothermal Research},
volume = {184},
number = {1},
pages = {193-198},
year = {2009},
note = {Recent advances on the geodynamics of Piton de la Fournaise volcano},
issn = {0377-0273},
doi = {https://doi.org/10.1016/j.jvolgeores.2008.10.002},
url = {https://www.sciencedirect.com/science/article/pii/S0377027308005428},
author = {S. Deroussi and M. Diament and J.B. Feret and T. Nebut and Th. Staudacher}
}

@article{Urbancic:2017,
author = {Urbancic, N. and Ghent, R. and Johnson, C. L. and Stanley, S. and Hatch, D. and Carroll, K. A. and Garry, W. B. and Talwani, M.},
title = {Subsurface density structure of Taurus-Littrow Valley using Apollo 17 gravity data},
journal = {Journal of Geophysical Research: Planets},
volume = {122},
number = {6},
pages = {1181-1194},
doi = {https://doi.org/10.1002/2017JE005296},
url = {https://agupubs.onlinelibrary.wiley.com/doi/abs/10.1002/2017JE005296},
eprint = {https://agupubs.onlinelibrary.wiley.com/doi/pdf/10.1002/2017JE005296},
year = {2017}
}

@article{Talwani:1976,
  title={Apollo 17 traverse gravimeter experiment/preliminary results},
  journal ={Geologisches Jahrbuch, Reihe E - Geophysik},
  pages ={85-91},
  author={Talwani, M and Kahle, H-G},
  year={1976}
}

@inproceedings{Buck:1973,
  title={Traverse gravimeter experiment},
  author={Buck, S},
  booktitle={Guidance and Control Conference},
  pages={1100},
  year={1973}
}

@article{Michel:2022,
  title={The ESA Hera mission: detailed characterization of the DART impact outcome and of the binary asteroid (65803) Didymos},
  author={Michel, Patrick and K{\"u}ppers, Michael and Bagatin, Adriano Campo and Carry, Benoit and Charnoz, S{\'e}bastien and De Leon, Julia and Fitzsimmons, Alan and Gordo, Paulo and Green, Simon F and H{\'e}rique, Alain and others},
  journal={The planetary science journal},
  volume={3},
  number={7},
  pages={160},
  year={2022},
  publisher={IOP Publishing}
}

@article{cushing2007,
author = {Cushing, G. E. and Titus, T. N. and Wynne, J. J. and Christensen, P. R.},
title = {THEMIS observes possible cave skylights on Mars},
journal = {Geophysical Research Letters},
volume = {34},
number = {17},
pages = {},
doi = {https://doi.org/10.1029/2007GL030709},
url = {https://agupubs.onlinelibrary.wiley.com/doi/abs/10.1029/2007GL030709},
eprint = {https://agupubs.onlinelibrary.wiley.com/doi/pdf/10.1029/2007GL030709},
year = {2007}
}

@article{hoek_practical_1997,
	title = {Practical estimates of rock mass strength},
	volume = {34},
    year ={1997},
	issn = {1365-1609},
	url = {https://www.sciencedirect.com/science/article/pii/S136516099780069X},
	doi = {10.1016/S1365-1609(97)80069-X},
	pages = {1165--1186},
	number = {8},
	journal = {International Journal of Rock Mechanics and Mining Sciences},
	author = {Hoek, E. and Brown, E. T.},
	urldate = {2026-02-20},
	date = {1997-12-01},
}

@article{roche_sub-surface_2001,
	title = {Sub-surface structures and collapse mechanisms of summit pit craters},
	volume = {105},
    year ={2001},
	issn = {0377-0273},
	url = {https://www.sciencedirect.com/science/article/pii/S0377027300002481},
	doi = {10.1016/S0377-0273(00)00248-1},
	pages = {1--18},
	number = {1},
	journal = {Journal of Volcanology and Geothermal Research},
	author = {Roche, O. and van Wyk de Vries, B. and Druitt, T. H.},
	urldate = {2026-02-03},
	date = {2001-01-01},
}

@article{okubo_pit_1998,
	title = {Pit crater formation on Kilauea volcano, Hawaii.},
	volume = {86},
	issn = {0377-0273},
    year ={1998},
	url = {https://ui.adsabs.harvard.edu/abs/1998JVGR...86....1O},
	doi = {10.1016/S0377-0273(98)00070-5},
	pages = {1--18},
	journal = {Journal of Volcanology and Geothermal Research},
	author = {Okubo, Chris H. and Martel, Stephen J.},
	urldate = {2026-02-03},
	date = {1998-11-01},
	note = {Publisher: Elsevier
{ADS} Bibcode: 1998JVGR...86....1O},
}

@article{Tomasi2022,
  author = {Tomasi, Isabella and Massironi, Matteo and Meyzen, Christine M. and Pozzobon, Riccardo and Sauro, Francesco and Penasa, Luca and Santagata, Tommaso and Tonello, Matteo and Santana Gomez, Gonzalo Daniel and Martinez-Frías, Jesús},
  title = {Inception and Evolution of La Corona Lava Tube System (Lanzarote, Canary Islands, Spain)},
  journal = {Journal of Geophysical Research: Solid Earth},
  volume = {127},
  number = {6},
  pages = {e2022JB024056},
  year = {2022},
  doi = {10.1029/2022JB024056}
}

@article{zunino2022hamiltonian,
  title={Hamiltonian Monte Carlo probabilistic joint inversion of 2D (2.75 D) gravity and magnetic data},
  author={Zunino, Andrea and Ghirotto, Alessandro and Armadillo, Egidio and Fichtner, Andreas},
  journal={Geophysical Research Letters},
  volume={49},
  number={20},
  pages={e2022GL099789},
  year={2022},
  publisher={Wiley Online Library}
}

@incollection{ghirotto2026seismic,
  title={Seismic, magnetic and gravity investigations of Lunar lava tubes: An Earth-analogue case study from Lanzarote island (Spain)},
  author={Ghirotto, Alessandro and Barone, Ilaria and Santoro De Vico, Francesco and Melchiori, Giacomo and Zunino, Andrea and Armadillo, Egidio and Mittelholz, Anna and Sauro, Francesco and Massironi, Matteo},
  booktitle={EGU General Assembly 2026 Abstracts},
  pages={1--1},
  year={2026}
}

@inproceedings{noeker2022grass,
  title={The GRASS gravimeter rotation mechanism for ESA hera mission on-board Juventas deep space CubeSat},
  author={Noeker, Matthias and Van Ransbeeck, Emiel and Ritter, Birgit and Karatekin, {\"O}zg{\"u}r},
  booktitle={Proceedings of the 46th Aerospace Mechanisms Symposium, Virtual},
  pages={159--172},
  year={2022}
}

@article{martellato_numerical_2013,
	title = {Numerical modelling of impact crater formation associated with isolated lunar skylight candidates on lava tubes},
	volume = {86},
	issn = {0032-0633},
	url = {https://www.sciencedirect.com/science/article/pii/S0032063313001517},
	doi = {10.1016/j.pss.2013.06.010},
	urldate = {2025-12-10},
	journal = {Planetary and Space Science},
	author = {Martellato, E. and Foing, B. H. and Benkhoff, J.},
	month = sep,
	year = {2013},
	pages = {33--44},
}

@article{blair_structural_2017,
	title = {The structural stability of lunar lava tubes},
	volume = {282},
	issn = {0019-1035},
	url = {https://www.sciencedirect.com/science/article/pii/S0019103516303566},
	doi = {10.1016/j.icarus.2016.10.008},
	urldate = {2025-08-18},
	journal = {Icarus},
	author = {Blair, David M. and Chappaz, Loic and Sood, Rohan and Milbury, Colleen and Bobet, Antonio and Melosh, H. Jay and Howell, Kathleen C. and Freed, Andrew M.},
	month = jan,
	year = {2017},
	pages = {47--55},
}

@article{Hamran:2020,
    author = {Hamran, Svein-Erik and Paige, David and Amundsen, Hans and et al.},
    doi = {10.1007/s11214-020-00740-4},
    journal = {Space Sci Rev},
    number = {216},
    pages = {1--39},
    title = {{ Radar Imager for Mars’ Subsurface Experiment—RIMFAX }},
    volume = {216},
    year = {2020}
}

@article{Qian2021,
author = {Qian, Yuqi and Xiao, Long and Head, James W. and Wilson, Lionel},
title = {The Long Sinuous Rille System in Northern Oceanus Procellarum and Its Relation to the Chang'e-5 Returned Samples},
journal = {Geophysical Research Letters},
volume = {48},
number = {11},
pages = {e2021GL092663},
doi = {https://doi.org/10.1029/2021GL092663},
url = {https://agupubs.onlinelibrary.wiley.com/doi/abs/10.1029/2021GL092663},
eprint = {https://agupubs.onlinelibrary.wiley.com/doi/pdf/10.1029/2021GL092663},
note = {e2021GL092663 2021GL092663},
year = {2021}
}

@article{HENRIKSEN2017,
title = {Extracting accurate and precise topography from LROC narrow angle camera stereo observations},
journal = {Icarus},
volume = {283},
pages = {122-137},
year = {2017},
note = {Lunar Reconnaissance Orbiter - Part II},
issn = {0019-1035},
doi = {https://doi.org/10.1016/j.icarus.2016.05.012},
url = {https://www.sciencedirect.com/science/article/pii/S001910351630152X},
author = {M.R. Henriksen and M.R. Manheim and K.N. Burns and P. Seymour and E.J. Speyerer and A. Deran and A.K. Boyd and E. Howington-Kraus and M.R. Rosiek and B.A. Archinal and M.S. Robinson}
}

@incollection{Haviland2025,
  title={Probing the lunar interior with electromagnetic geophysical methods},
  author={Haviland, Heidi Fuqua and Mittelholz, Anna},
  booktitle={Advances in Geophysics},
  volume={66},
  pages={135--177},
  year={2025},
  publisher={Elsevier}
}

@book{center_us_apollo_1972,
    title = {Apollo 16: {Preliminary} {Science} {Report}},
    shorttitle = {Apollo 16},
    language = {en},
    publisher = {Scientific and Technical Information Office, National Aeronautics and Space Administration},
    author = {Center (U.S.), Manned Spacecraft},
    year = {1972},
    note = {Google-Books-ID: yI41Km4\_CDUC},
}
\end{document}